# Transition Metal Dichalcogenide Solar Cells for Indoor Energy Harvesting

Frederick U. Nitta,[1,2] Koosha Nassiri Nazif,[1] and Eric Pop,[1,2,3*]

[1]Dept. of Electrical Engineering, [2]Dept. of Materials Science and Engineering, [3]Dept. of Applied Physics, Stanford University, Stanford, CA 94305, USA. *Corresponding author email: epop@stanford.edu

*Abstract* — **With the rapid expansion of the Internet of Things (IoT), efficient and durable energy harvesters for powering IoT devices operating indoors and outdoors are imperative. Promising materials for indoor photovoltaic (PV) technologies include transition metal dichalcogenides (TMDs) such as $MoS_2$, $MoSe_2$, $WS_2$, and $WSe_2$, mainly due to their high absorption coefficients and self-passivated surfaces. Here, we assess the performance of single-junction TMD solar cells under various indoor lighting conditions with a realistic detailed balance model including material-specific optical absorption, as well as radiative, Auger, and defect-assisted Shockley-Read-Hall recombination. We find TMD solar cells could achieve up to 36.5%, 35.6%, 11.2%, and 27.6% power conversion efficiency under fluorescent, LED, halogen, and low-light AM 1.5 G lighting, respectively, at 500 lux. Based on this, TMD solar cells could outperform commercial PV technologies in indoor scenarios, suggesting their viability for future IoT energy solutions.**

## I. Introduction

As the Internet of Things (IoT) expands, the need for reliable energy sources to power IoT devices becomes increasingly vital, especially within indoor environments. Indoor photovoltaics (PVs) offer a sustainable solution, addressing the energy requirements for the vast network of sensors and devices that will form the backbone of data-driven sectors such as healthcare, manufacturing, infrastructure, and energy. It is anticipated that billions of wireless sensors will be deployed in the next decade, with a substantial number located indoors to facilitate continuous data acquisition and system optimization[1,2].

While several indoor PV technologies such as amorphous silicon (a-Si)[3], dye-sensitized solar cells (DSSC)[4,5], organic PV[6,7], and perovskite solar cells (PSC)[8–10] have been explored, each presents some challenges in terms of efficiency, stability, and production scalability[11–18]. Among emerging materials and technologies for indoor PV solutions, transition metal dichalcogenides (TMDs) are attracting attention due to their high absorption coefficients, near-ideal band gap, and self-passivated surfaces[19,20]. Models show that ultrathin TMD solar cells (~50 nm) can achieve 25% power conversion efficiency outdoors – under the AM 1.5 G spectrum – upon design optimization even with existing material quality. This corresponds to 10× higher specific power that of existing incumbent solar cell technologies[21]. Although similarly high performance is expected from TMD solar cells indoors, there are no prior studies of indoor performance of TMD solar cells that quantify the power output of TMD solar cells indoors.

In this work, we provide material-specific, thickness-dependent efficiency limits for single-junction solar cells made of multilayer (bulk, ≥ 5 nm-thick) $MoS_2$, $MoSe_2$, $WS_2$, and $WSe_2$ solar cells at different material qualities and under various indoor lighting conditions. We use a realistic detailed balance model incorporating experimental optical properties as well as radiative, Auger, and Shockley-Read-Hall recombination mechanisms. The performance of these solar cells is analyzed under various indoor light sources, including compact fluorescent lamp (CFL), light emitting diode (LED), halogen, and low-intensity AM 1.5 G, all adjusted to the illuminance levels typical in common indoor locations ranging from parking garages (50 lux) to retail stores (500 lux). We find that TMD solar cells could achieve high power conversion efficiencies – up to 36.5% under CFL, 35.6% under LED, 11.2% under halogen lighting, and 27.6% under low-light AM 1.5 G. This performance indicates that TMDs offer an improvement over existing indoor PV technologies, potentially transforming energy solutions for indoor IoT applications.

## II. RESULTS AND DISCUSSION

### Modeling setup

Our modeling approach, detailed in our previous work[21] and in **Supplementary Note 1**, extends beyond the Tiedje-Yablonovitch limit[22] to investigate the impact of material quality on solar cell performance. It incorporates defect-assisted Shockley-Read-Hall (SRH) recombination to establish efficiency limits for single-junction, multilayer TMD solar cells with film thicknesses of 5 nm or more as a function of material quality. It considers an enhanced absorption via a mean path length of $4n^2L$ (with $n$ representing the refractive index), and photogenerated excitons that immediately dissociate into free charge carriers (**Figure 1a**).

We examined the efficiency limits of TMD solar cells under four indoor spectra[23]: compact fluorescent lamp (CFL) also known as energy saving lamp (ESL), incandescent halogen, light-emitting diode (LED), and low-light AM 1.5 G. These spectra are all shown in **Figure 1b**. The halogen spectrum was extended using the blackbody radiation formula (**Supplementary Note 2** and **Supplementary Figure 1**) to match the halogen lamp's emission characteristics. Normalization of these spectra to typical indoor lighting scenarios was achieved by matching the lux levels defined in the Illuminating Engineering Society (IES) Lighting Handbook[24]. These scenarios range from the lower intensity of a parking lot at 50 lux and a warehouse at 150 lux, to brighter conditions of an office at 400 lux and a retail store at 500 lux.

**Figure 1b** shows the normalization of these spectra for a retail setting (500 lux) as an example. The lux illumination was calculated by calibrating the spectral power distribution with the Commission Internationale de l'Eclairage (CIE) photopic luminosity function. Although lux values are based on the human-visible range, the input power calculations consider the full spectra of light wavelengths, which are needed to determine the power conversion efficiency (the ratio of output power to input power). The input power densities for the four indoor spectra at the lux levels considered are listed in **Supplementary Table 1**. This calibration process is in line with methodologies applied to the AM 1.5 G spectrum in a previous study[25].

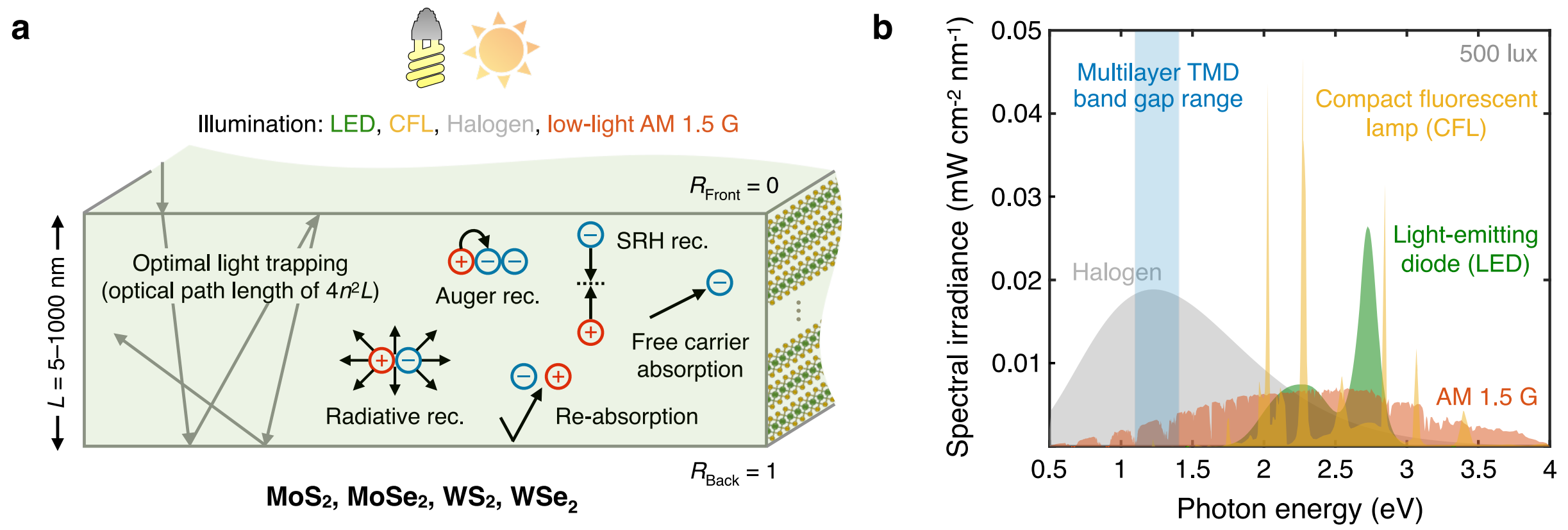

**Figure 1. Indoor TMD photovoltaics. a**, Modeling setup showing the configuration of a multilayer transition metal dichalcogenide (TMD) solar cell with various illumination sources. Optimal light trapping with an optical path length of *4n²L* is considered and various recombination mechanisms are included. Anti-reflection coatings and a perfect back reflector are used to enhance light absorption and minimize reflective losses. *R*, reflection; *L*, TMD film thickness; *n*, refractive index; SRH rec., Shockley-Read-Hall recombination. **b**, Various indoor light spectra[23] used in this study. The band gap range of multilayer TMDs is labeled by the blue shaded region. All spectra are normalized at 500 lux (retail store condition).

The following sections delve into a detailed analysis of the results, including short-circuit current density, open-circuit voltage, fill factor, output power, and power conversion efficiency of single-junction TMD ($MoS_2$, $MoSe_2$, $WS_2$, and $WSe_2$) solar cells for each considered light source (CFL, LED, halogen, and AM 1.5 G) at various illuminance levels typical in common indoor locations, ranging from parking garages (50 lux) to retail stores (500 lux).

**Compact fluorescent lamp (CFL)**

To illustrate the CFL estimates for one of the TMDs (here, $WS_2$), **Figure 2** shows the calculated short-circuit current density ($J_{SC}$), open-circuit voltage ($V_{OC}$), fill factor (FF), and output power ($P_{out}$) under CFL lighting as a function of $WS_2$ film thickness and CFL illumination intensity. For this study, our choice of $\tau_{SRH}$ = 611 ns is based on the maximum value reported to date for unpassivated multilayer $WS_2$[26]; our expectation is that as material quality continues to improve (and/or TMD surfaces are passivated), the lifetimes for $WS_2$ and other TMDs will increase beyond this figure. In comparison, an infinite SRH lifetime represents an idealized scenario (the Tiedje-Yablonovitch limit[22]), which points to the maximum achievable $V_{OC}$ (and efficiency) in the absence of defect-assisted SRH recombination.

The $J_{SC}$ in **Figure 2a** has minimal variation with increasing film thickness at low light intensities, such as those in parking garages (50 lux) or warehouses (150 lux). However, at the higher intensities in office (400 lux) and retail (500 lux) environments, the enhanced absorption of lower-energy photons by thicker $WS_2$ films leads to a modest rise in $J_{SC}$. **Figure 2b** finds the $V_{OC}$ similarly increases with light intensity, but decreases with thicker films, due to a shift in absorption threshold towards lower photon energies[21], which effectively reduces the band gap. The scaling of $V_{OC}$ with light intensity stems from the logarithmic relationship between $V_{OC}$ and photocurrent. The FF is influenced by both the $V_{OC}$ and the material quality. For a finite SRH lifetime of 611 ns, both the $V_{OC}$ and the FF (**Figure 2c**) are lower, and FF shows a stronger dependence on light intensity due to the greater relative impact of recombination at defect sites under lower light conditions. Although recombination is reduced at lower light intensities, the fewer available carriers make any recombination losses more detrimental to the FF, particularly in thicker films with more defects. The dependency of the FF on $V_{OC}$ further explains its reduction with increasing film thickness[27]. Lastly, reflecting trends from the other parameters, the output power $P_{out}$ (**Figure 2d)** displays a weak inverted U-shape as a function of film thickness. The $P_{out}$ peaks at intermediate thicknesses where the increase in $J_{SC}$, particularly at the higher light intensities, compensates for losses in $V_{OC}$ and FF.

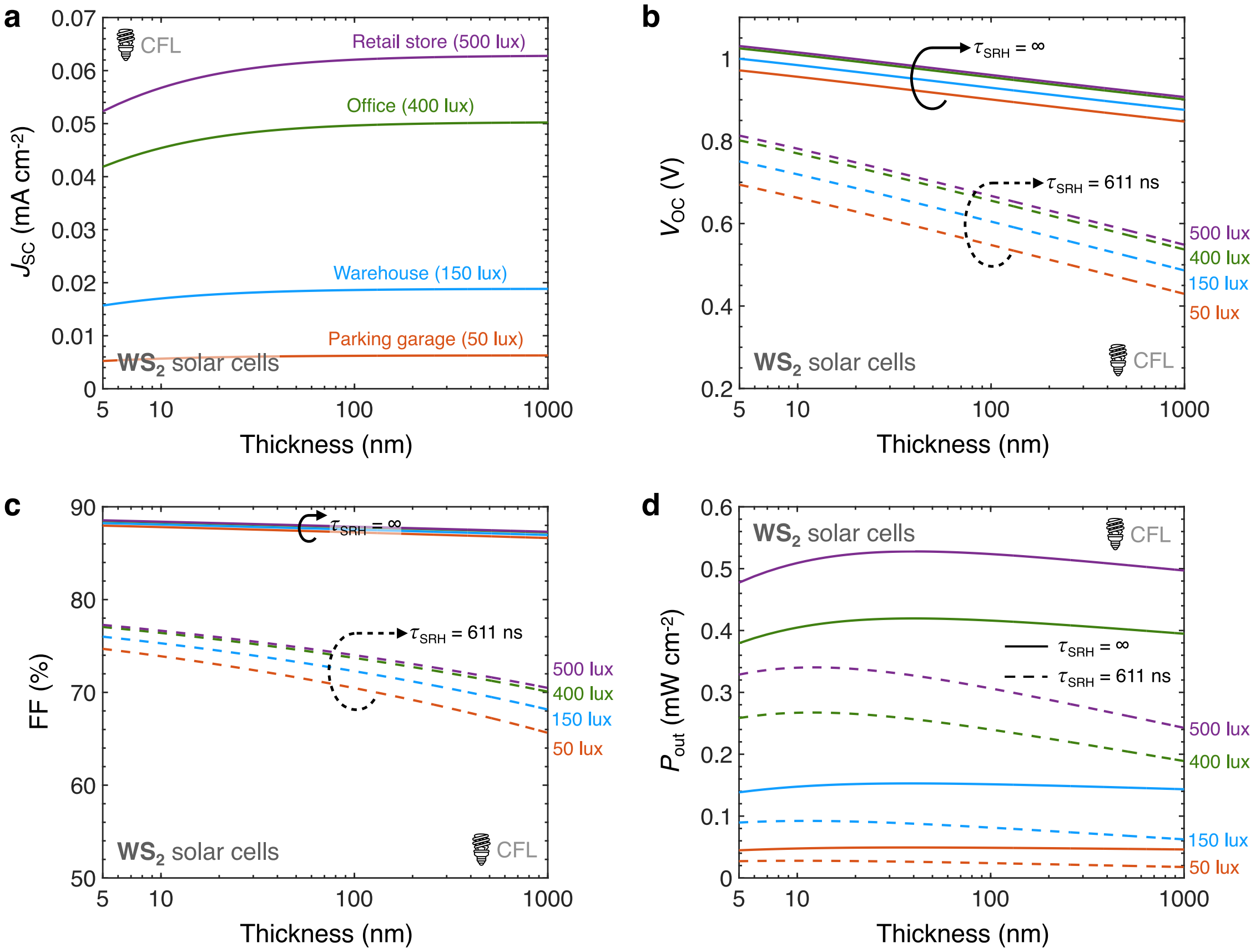

**Figure 2. $WS_2$ solar cells with compact fluorescent lamp (CFL) illumination. a**, $J_{SC}$, **b**, $V_{OC}$, **c**, FF, and **d**, $P_{out}$, all as a function of $WS_2$ film thickness, at 300 K. Solid lines are in the limit defect-free material (no SRH recombination), dashed lines with $\tau_{SRH}$ = 611 ns. Four CFL illumination intensities correspond to the four colors, as labeled (e.g. purple dashed and solid are at 500 lux). Note, $J_{SC}$ is not affected by material quality ($\tau_{SRH}$) due to the low carrier density at zero bias.

The power conversion efficiency (PCE) for $MoS_2$, $MoSe_2$, $WS_2$, and $WSe_2$ solar cells as a function of TMD film thickness, SRH lifetime ($\tau_{SRH}$), and CFL illumination intensity is shown in **Figure 3**. PCE is the ratio of the $P_{out}$ to the input power ($P_{in}$), and it characterizes the efficiency with which the solar cells convert the absorbed light into electrical power. Because the $P_{out}$ of solar cells is the product of $J_{SC}$, $V_{OC}$, and FF, these trends are explained by the $J_{SC}$, $V_{OC}$, and FF trends in **Supplementary Figure 2**, **Supplementary Figure 3**, and **Supplementary Figure 4**, respectively. As observed, the $P_{out}$ curves for all four TMDs (**Supplementary Figure 5**) exhibit an inverted U-shape, which also defines the PCE curve due to the competing influences of $J_{SC}$, $V_{OC}$, and FF on the $P_{out}$. As the TMD film thickness increases, the $J_{SC}$ improves due to better light absorption, but both $V_{OC}$ and FF decrease, which is more pronounced for the finite SRH lifetime of 611 ns. This competition leads to the same trends observed for PCE as for $P_{out}$ with thickness.

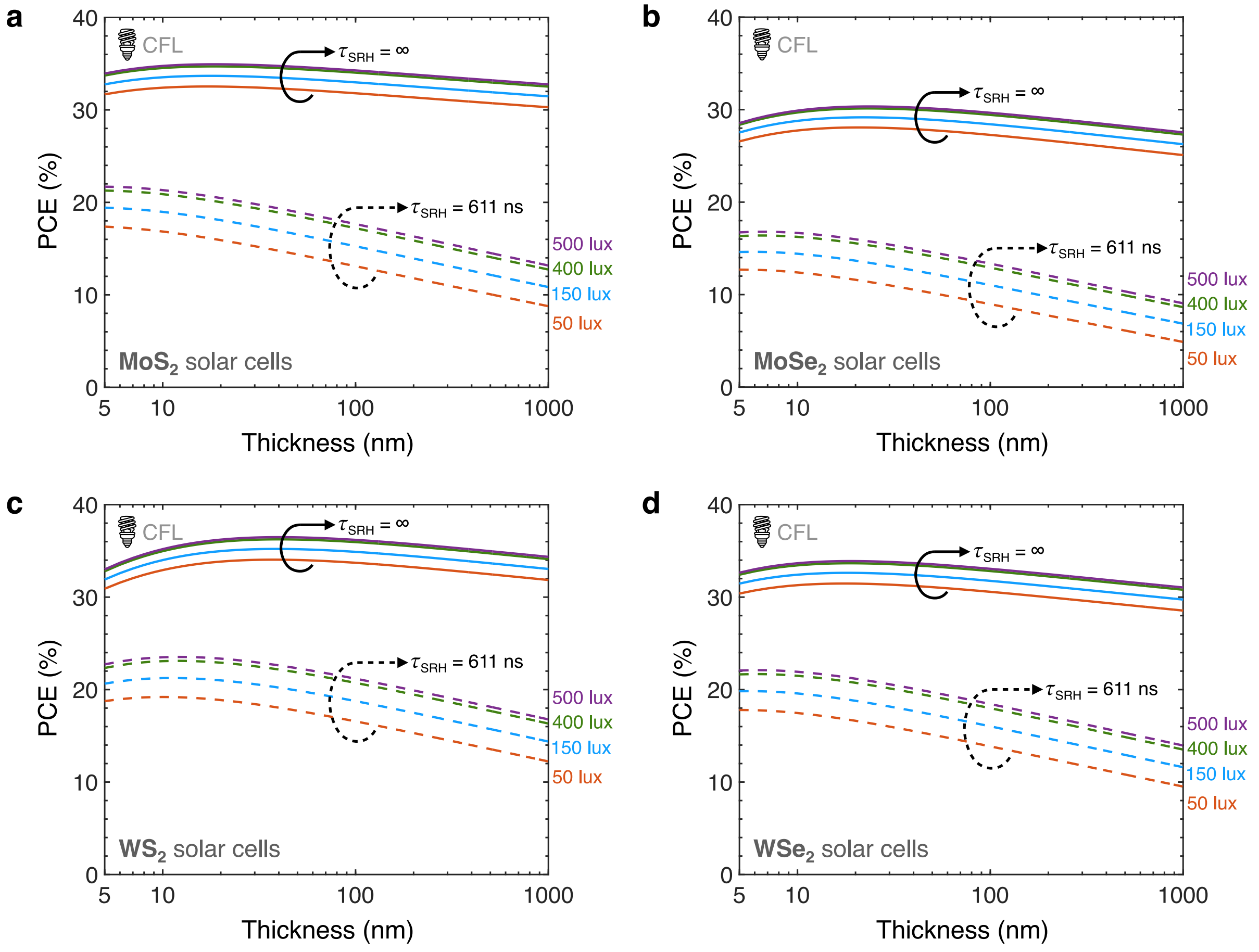

**Figure 3. Power conversion efficiency (PCE) of thin-film TMD solar cells under CFL illumination.** PCE of **a,** $MoS_2$, **b,** $MoSe_2$, **c,** $WS_2$, and **d,** $WSe_2$ solar cells as a function of TMD film thickness, material quality ($\tau_{SRH}$), and CFL illumination intensity at 300 K. Solid lines are in the limit defect-free material (no SRH recombination), dashed lines with $\tau_{SRH}$ = 611 ns. Illumination intensities correspond to the four colors, as labeled (e.g. purple dashed and solid line are at 500 lux). $\tau_{SRH}$, Shockley-Read-Hall (SRH) lifetime.

For lower light intensities, even though the $P_{out}$ may not exhibit a distinct peak (**Supplementary Figure 5**), the division by a relatively smaller input power accentuates the peak in PCE (**Figure 3**). For infinite SRH lifetime where non-radiative recombination is excluded, the degradation in PCE with increased thickness is less severe, illustrating the critical role of material quality in TMD solar cell performance. Conversely, in the presence of SRH recombination, due to the steeper decline in $V_{OC}$ and FF with thickness (**Supplementary Figure 3** and **Supplementary Figure 4**), the PCE exhibits a peak shift towards smaller thicknesses, as well as a more significant drop-off with thickness.

With today's material quality ($\tau_{SRH} \approx 611$ ns), TMD solar cells can achieve up to 23.5% efficiency under CFL illumination. The efficiency limits at current material quality could be achieved through careful optimization of the solar cell's optical and electrical designs. As material quality improves towards an infinite $\tau_{SRH}$, efficiencies as high as 36.5% become achievable, underscoring that better material quality directly correlates with enhanced performance.

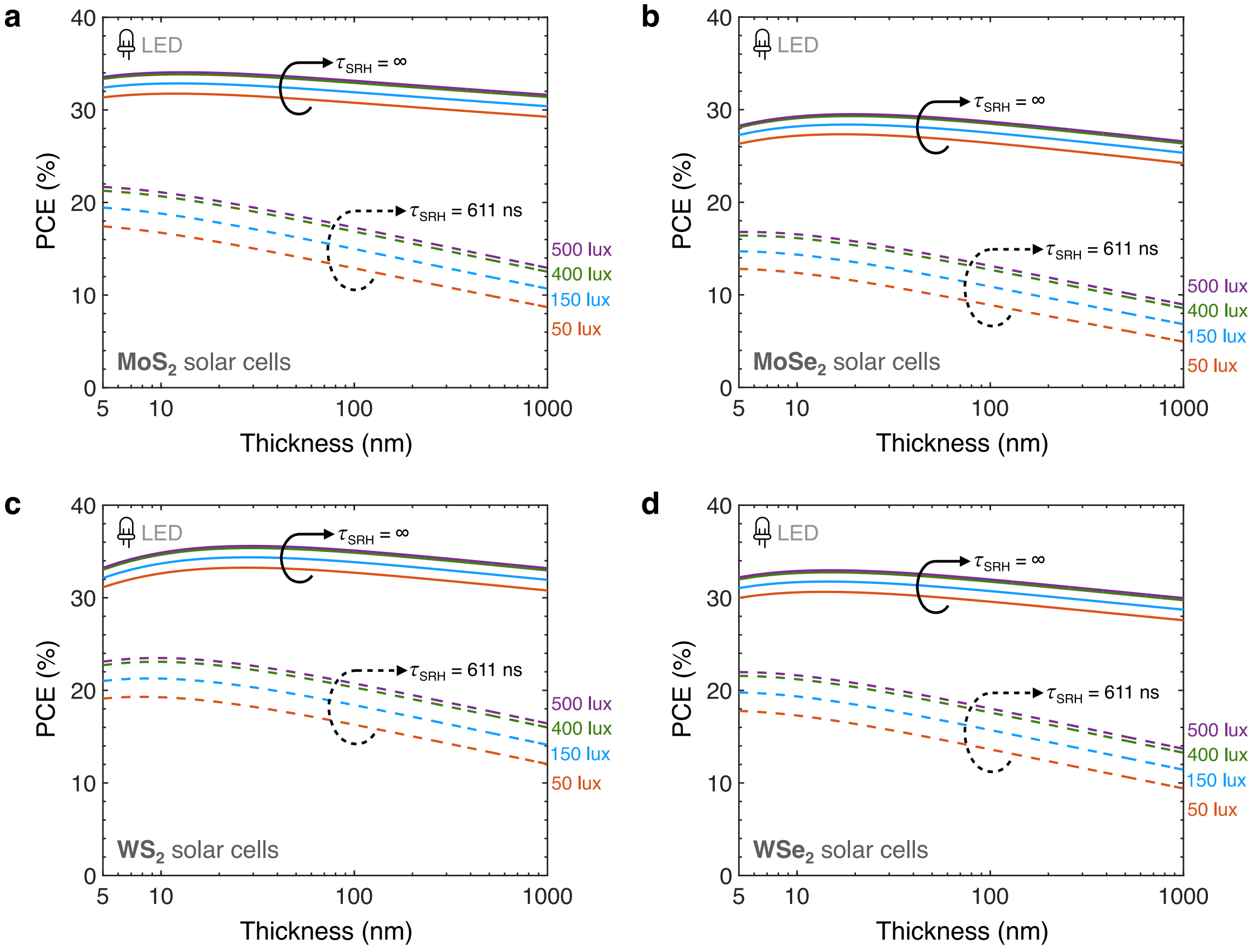

**Figure 4. Power conversion efficiency (PCE) of thin-film TMD solar cells under LED illumination.** PCE of **a,** $MoS_2$, **b,** $MoSe_2$, **c,** $WS_2$, and **d,** $WSe_2$ solar cells as a function of TMD film thickness, material quality ($\tau_{SRH}$), and LED illumination intensity at 300 K. Solid lines are in the limit defect-free material (no SRH recombination), dashed lines with $\tau_{SRH}$ = 611 ns. Illumination intensities correspond to the four colors, as labeled (e.g. purple dashed and solid line are at 500 lux). $\tau_{SRH}$, Shockley-Read-Hall (SRH) lifetime.

## Light-emitting diode (LED) lamp

We now examine the PCE for all four TMD solar cells with LED illumination, as a function of TMD film thickness, illumination intensity, and SRH lifetime ($\tau_{SRH}$), as shown in **Figure 4**. Thicker films initially enhance $J_{SC}$ (**Supplementary Figure 6**), contributing to an increase in PCE due to improved light absorption capabilities. This increase in PCE is, however, countered by decreases in $V_{OC}$ (**Supplementary Figure**

7) and FF (**Supplementary Figure 8**), particularly where the SRH lifetime is finite. The $P_{out}$ trends (**Supplementary Figure 9**), and thus the PCE trends, similar to those under CFL illumination, show that there is an optimal thickness where benefits in $J_{SC}$ are maximized before being outweighed by losses in $V_{OC}$ and FF. With infinite SRH lifetime, the drop in PCE with increased thickness is not as pronounced, thanks to the higher material quality. At finite SRH lifetimes, a sharper peak and a more noticeable decline in PCE with additional thickness are observed. These trends suggest that enhancing material quality can significantly improve the power conversion efficiency of TMD-based solar cells under LED lighting.

With today's material quality ($\tau_{SRH} \approxeq 611$ ns), TMD solar cells can achieve up to 23.5% efficiency under LED illumination. As material quality advances towards infinite $\tau_{SRH}$, there is potential to reach efficiencies as high as 35.6%. These enhancements in efficiency can be realistically attained by optimizing the optical and electrical designs of the TMD solar cells, and leveraging improvements in material quality.

### Halogen lamp

Examining the four TMD solar cells with halogen illumination, **Figure 5** displays their estimated PCE as a function of TMD film thickness, SRH lifetime ($\tau_{SRH}$), and illumination intensity. The trends in $J_{SC}$ (**Supplementary Figure 10**), $V_{OC}$ (**Supplementary Figure 11**), and FF (**Supplementary Figure 12**) inform the trends in $P_{out}$ (**Supplementary Figure 13**) and thus PCE. Notably, for an infinite SRH lifetime, we observe a continuous increase in PCE across all materials, suggesting the outweighing benefit of improvements in $J_{SC}$ with thicker films against losses in $V_{OC}$ and FF. However, the PCE under halogen illumination is overall lower than the PCE under CFL or LED illumination (**Figures 3** and **4**), due to the halogen spectrum having more low-energy photons, below the TMD band gaps (**Figure 1b**). For a finite $\tau_{SRH}$ of 611 ns, $MoS_2$ and $WSe_2$ exhibit discernible PCE peaks, indicating an optimal thickness range for maximum efficiency. In contrast, the PCE of $MoSe_2$ (**Figure 5b**) is almost independent of film thickness, which points to its less dramatic balance between $J_{SC}$ gains and $V_{OC}$ and FF losses. The PCE of $WS_2$ (**Figure 5c**) shows a monotonic increase with film thickness, which is more pronounced at the higher illumination intensities. This demonstrates how the material's absorption spectrum and the illumination intensity affects PCE.

With today's material quality, at $\tau_{SRH} \approx 611$ ns (dashed lines in **Figure 5**), these TMD solar cells can achieve up to 5.9% efficiency under halogen illumination. As material quality advances towards infinite $\tau_{SRH}$ (solid lines in **Figure 5**) efficiencies up to 11.2% may be reachable in the thickest $MoS_2$ and $WSe_2$ films, under halogen illumination.

### Low-Light AM 1.5 G

The PCEs for $MoS_2$, $MoSe_2$, $WS_2$, and $WSe_2$ solar cells as a function of TMD film thickness, SRH lifetime ($\tau_{SRH}$), and AM 1.5 G illumination intensity are shown in **Figure 6**. Similar to observations under halogen illumination, for an infinite SRH lifetime, the PCEs consistently increase with film thickness for all materials, indicating that the positive effects of increased $J_{SC}$ (**Supplementary Figure 14**) with thickness outweigh the negative impacts on $V_{OC}$ (**Supplementary Figure 15**) and FF (**Supplementary Figure 16**). Similar trends are seen for $P_{out}$ (**Supplementary Figure 17**) within infinite SRH lifetime.

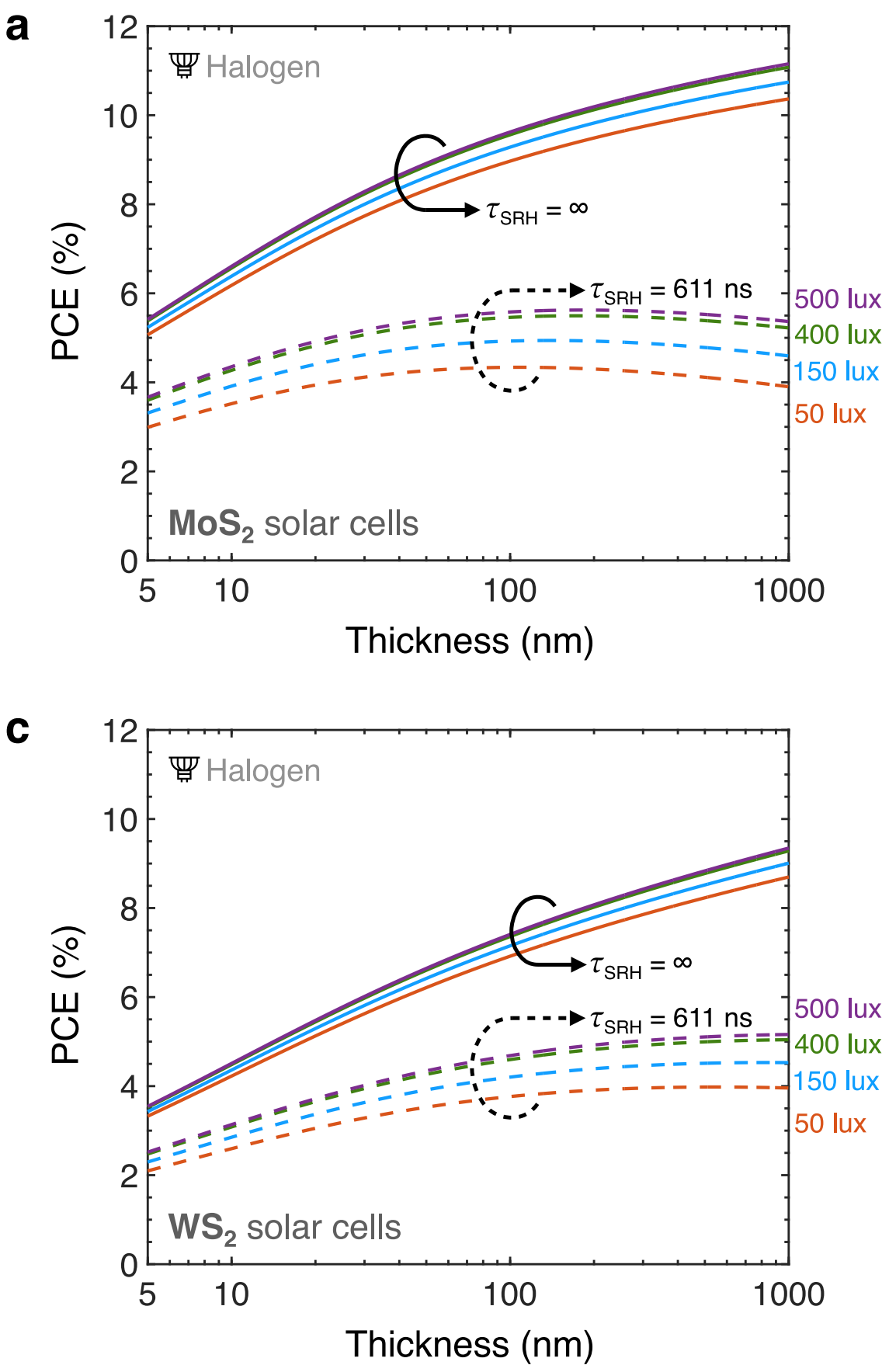

**Figure 5. Power conversion efficiency (PCE) of thin-film TMD solar cells under halogen illumination.** PCE of **a,** $MoS_2$, **b,** $MoSe_2$, **c,** $WS_2$, **d,** $WSe_2$ solar cells as a function of TMD film thickness, material quality ($\tau_{SRH}$), and halogen illumination intensity at 300 K. Solid lines are in the limit defect-free material (no SRH recombination), dashed lines with $\tau_{SRH}$ = 611 ns. Illumination intensities correspond to the four colors, as labeled (e.g. purple dashed and solid line are at 500 lux). $\tau_{SRH}$, Shockley-Read-Hall (SRH) lifetime.

However, for a finite SRH lifetime of 611 ns, $P_{out}$ and PCE curves exhibit distinct peaks for all materials, similar to trends seen with CFL and LED illumination. These peaks highlight the interplay between $J_{SC}$, $V_{OC}$, and FF in determining the efficiency of light absorption and conversion to electrical power. Increasing film thickness enhances $J_{SC}$ due to improved light absorption, but this benefit is counterbalanced by more substantial declines in both $V_{OC}$ and FF in the thicker films.

With today's material quality ($\tau_{SRH} \approx 611$ ns), TMD solar cells achieve up to a PCE of 16.3% under AM 1.5 G illumination, at 500 lux (i.e. typical retail store lighting). As material quality progresses towards an infinite SRH lifetime, the efficiency could increase to as much as 27.6%. These efficiency limits are attainable by refining the optical and electrical designs of these ultrathin TMD solar cells.

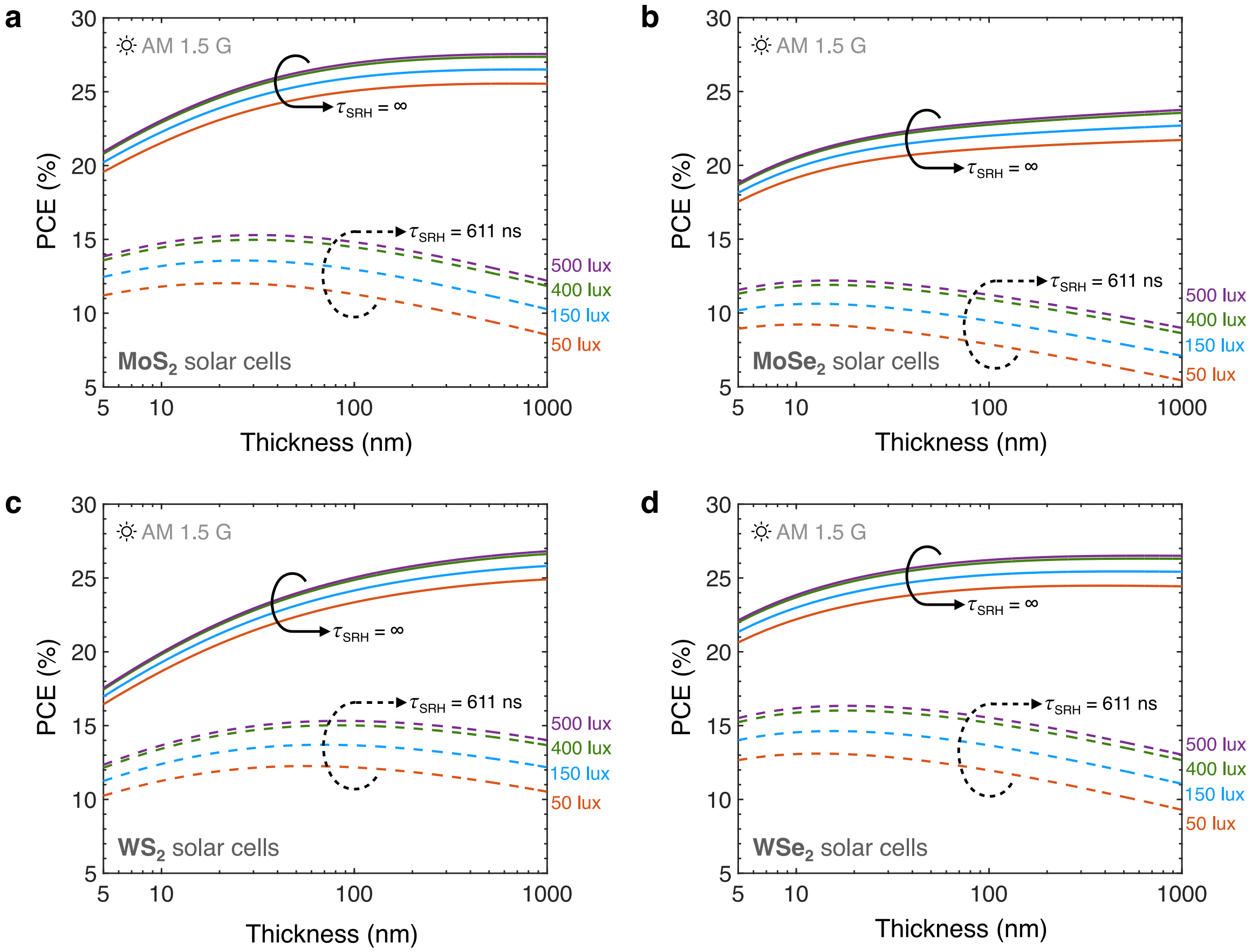

**Figure 6. Power conversion efficiency (PCE) of thin-film TMD solar cells under AM 1.5 G illumination.** PCE of **a,** $MoS_2$, **b,** $MoSe_2$, **c,** $WS_2$, **d,** $WSe_2$ solar cells as a function of TMD film thickness, material quality ($\tau_{SRH}$), and AM 1.5 G illumination intensity at 300 K. Solid lines are in the limit defect-free material (no SRH recombination), dashed lines with $\tau_{SRH}$ = 611 ns. Illumination intensities correspond to the four colors, as labeled (e.g. purple dashed and solid line are at 500 lux). $\tau_{SRH}$, Shockley-Read-Hall (SRH) lifetime.

## Benchmarking and Projections

**Figure 7a** benchmarks the PCE of TMD solar cells in this work against incumbent and emerging indoor photovoltaic technologies, under CFL and LED illumination at ~500 lux, as detailed in **Supplementary Table 2**. Shockley-Queisser efficiency limits considering a CFL spectrum and an LED spectrum at 500 lux are included for reference (solid lines). For CFL illumination, TMD solar cells achieve PCE up to 23.5% at $\tau_{SRH}$ of 611 ns, and 36.5% in the absence of SRH recombination, both for $WS_2$. Under LED illumination, these efficiencies are slightly lower at 23.5% and 35.6%, respectively. For an infinite $\tau_{SRH}$, TMD solar cell efficiencies above the Shockley-Queisser model, are due to our incorporation of measured optical absorption spectra, which show that the absorption threshold of TMDs shifts to higher photon energies in thinner films, effectively increasing the band gap. This results in higher open-circuit voltages ($V_{OC}$) than predicted by the Shockley-Queisser model, and thus higher FF, leading to higher overall PCEs[21].

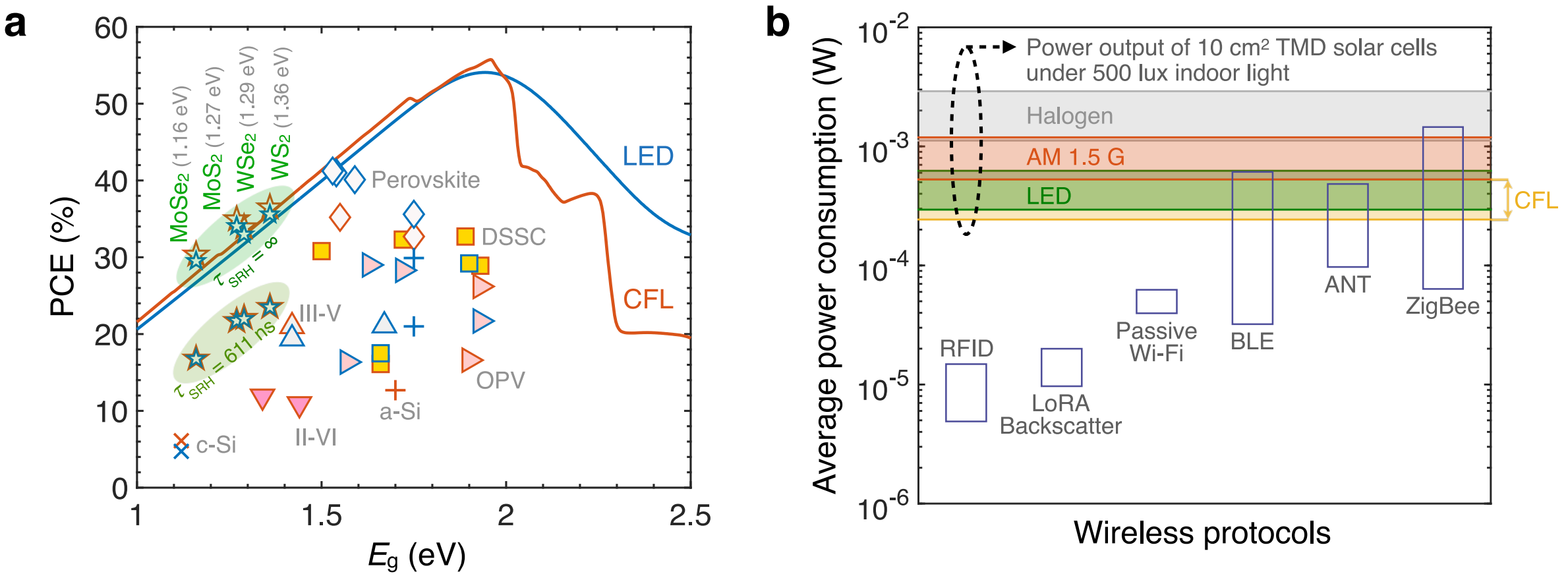

**Figure 7. Comparison of TMD solar cells with other indoor PV technologies and IoT wireless protocols. a,** Maximum power conversion efficiencies (PCE) vs. band gap ($E_g$) for indoor PV devices to date; symbols with orange (blue) border mark measurements under CFL (LED) illumination. Orange (blue) line marks the Shockley-Queisser limit at 500 lux with CFL (LED) illumination spectrum. Indoor PV PCEs are at approximately 500 lux to 1000 lux illumination; see **Supplementary Table 2** for more details and references. TMD estimates from this work (stars) are at 500 lux illumination, at $\tau_{SRH}$ of 611 ns and infinite $\tau_{SRH}$, for both CFL and LED. **b,** Comparison of average power consumption of various wireless protocols[2,28,29] with the $P_{out}$ of 10 cm$^2$ TMD solar cells under various indoor lighting conditions at 500 lux. This highlights the ability of TMD solar cells to efficiently power IoT devices across multiple indoor settings, underscoring their potential to support the sustainable expansion of IoT networks.

Our estimated TMD solar cell efficiencies, achievable with optimized optical and electronic designs, are comparable to those of existing indoor PV technologies, such as DSSC and organic PV, under similar conditions. Even at a $\tau_{SRH}$ of 611 ns, which corresponds to existing TMD material quality[26], TMD solar cells are competitive with a-Si, III-V, II-VI and c-Si (crystalline Si) technologies. Notably, the efficiencies of TMD solar cells with "infinite" $\tau_{SRH}$ closely approach or surpass the highest efficiencies reported (details and references in **Supplementary Table 2**), demonstrating TMDs' strong potential in indoor applications. Although perovskites at the moment achieve the highest efficiency records for indoor PVs, they suffer from stability issues, both in the dark and under illumination[30], and use materials that raise environmental and health concerns[18,31]. In contrast, TMD solar cells avoid these drawbacks, being stable and free from toxic elements like lead, making them a safer and more sustainable choice for indoor applications. Moreover, advancements in TMD growth and device fabrication in the nanoelectronics industry[20,32–35] enable low-cost mass production of TMD solar cells[36], rendering TMD solar cells an excellent candidate for indoor PVs.

**Figure 7b** compares the average power requirement of various IoT communications protocols and the range of power outputs from 10 cm$^2$ TMD solar cells under various indoor light spectra at 500 lux. This figure shows that the power output of these TMD solar cells is sufficient to support a range of low-power network protocols essential for IoT applications, such as RFID (Radio Frequency Identification), LoRA (Long Range) backscatter, passive Wi-Fi, BLE (Bluetooth Low Energy), ANT, and Zigbee. These protocols are critical for ensuring efficient, reliable data transmission within indoor IoT networks. By meeting and potentially exceeding the power demands of current IoT protocols, TMD solar cells could be a key component in advancing sustainable IoT infrastructures.

## IV. Conclusions

We examined the potential of TMD solar cells for indoor energy harvesting, specifically assessing the performance of $MoS_2$, $MoSe_2$, $WS_2$, and $WSe_2$ of various thicknesses and under various indoor lighting conditions, including CFL, LED, halogen lighting, and low-light AM 1.5 G. Our detailed balance model incorporates material-specific optical absorption data and accounts for various recombination mechanisms, including radiative, Auger, and SRH processes. We find that TMD solar cells can outperform existing indoor PV technologies, with power conversion efficiency limits up to 36.5% under fluorescent lighting, 35.6% under LED, 11.2% under halogen lighting, and 27.6% under low-light AM 1.5 G, at 500 lux. With today's material quality, TMD solar cells can achieve up to 23.5% under fluorescent lighting, 23.5% under LED, 5.9% under halogen lighting, and 16.3% under low-light AM 1.5 G, at 500 lux. These efficiencies suggest the viability of TMD solar cells for powering IoT devices within indoor environments. Future work will need to focus on further refining the electrical and optical designs of TMD solar cells to fully capitalize on their high-efficiency potential and adapt them for broader commercial applications.

**Methods:** The detailed balance equation governing the current density–voltage characteristics of the solar cell and the method to extract the performance metrics, i.e. short-circuit current density, open-circuit voltage, fill factor, and power conversion efficiency is explained in detail in **Supplementary Note 1**. The code developed to solve the detailed balance equation is provided in the **Code Availability** section.

**Data Availability:** The data that support the findings of this study are available from the corresponding author upon reasonable request.

**Code Availability:** Code to replicate the main findings of this study can be found on GitHub at https://github.com/fnitta/indoor-photovoltaics-efficiency-limit.

**Authors Contributions:** K.N. conceived the project. F.U.N. and K.N. developed the extended detailed balance model. F.U.N. implemented the model on TMDs, assisted by K.N. All authors, i.e. F.U.N., K.N., and E.P., contributed to the data interpretation, presentation, and writing of the manuscript. E.P. supervised the work.

**Acknowledgements:** The authors acknowledge partial support from Stanford Precourt Institute for Energy and the member companies of the SystemX Alliance at Stanford.

**Competing Interests:** The authors declare no competing interests.

## REFERENCES

1. Ramamurthy, S. Wireless sensors: Technologies and global markets. *BCC Research, September* (2016).
2. Mathews, I., Kantareddy, S. N., Buonassisi, T. & Peters, I. M. Technology and market perspective for indoor photovoltaic cells. *Joule* 3, 1415–1426 (2019).
3. Yang, C. H. *et al.* Hydrogenated amorphous silicon solar cells on textured flexible substrate copied from a textured glass substrate template. *IEEE Electron Device Letters* 32, 1254–1256 (2011).
4. Chiba, Y. *et al.* Dye-sensitized solar cells with conversion efficiency of 11.1%. *Japanese Journal of Applied Physics, Part 2: Letters* 45, (2006).
5. Michaels, H. *et al.* Dye-sensitized solar cells under ambient light powering machine learning: Towards autonomous smart sensors for the internet of things. *Chem Sci* 11, 2895–2906 (2020).
6. Abdulrazzaq, O. A., Saini, V., Bourdo, S., Dervishi, E. & Biris, A. S. Organic solar cells: A review of materials, limitations, and possibilities for improvement. *Particulate Science and Technology* 31, 427–442 (2013).
7. Kang, H. *et al.* Bulk-Heterojunction Organic Solar Cells: Five Core Technologies for Their Commercialization. *Advanced Materials* 28, 7821–7861 (2016).
8. Li, Z. *et al.* Minimized surface deficiency on wide-bandgap perovskite for efficient indoor photovoltaics. *Nano Energy* 78, 105377 (2020).
9. Gong, O. Y. *et al.* Van der Waals force-assisted heat-transfer engineering for overcoming limited efficiency of flexible perovskite solar cells. *ACS Energy Lett* 7, 2893–2903 (2022).
10. Ann, M. H. *et al.* Device design rules and operation principles of high-power perovskite solar cells for indoor applications. *Nano Energy* 68, 104321 (2020).
11. Shimizu, T. Staebler-Wronski effect in hydrogenated amorphous silicon and related alloy films. *Japanese Journal of Applied Physics, Part 1: Regular Papers and Short Notes and Review Papers* 43, 3257–3268 (2004).
12. Matsui, T. *et al.* High-efficiency amorphous silicon solar cells: Impact of deposition rate on metastability. *Appl Phys Lett* 106, 053901 (2015).
13. Moser, M., Wadsworth, A., Gasparini, N. & McCulloch, I. Challenges to the Success of Commercial Organic Photovoltaic Products. *Adv Energy Mater* 11, 2100056 (2021).
14. Xie, L. *et al.* Recent progress of organic photovoltaics for indoor energy harvesting. *Nano Energy* 82, 105770 (2021).
15. Cui, Y., Hong, L. & Hou, J. Organic Photovoltaic Cells for Indoor Applications: Opportunities and Challenges. *ACS Appl Mater Interfaces* 12, 38815–38828 (2020).
16. Duan, L. *et al.* Stability challenges for the commercialization of perovskite–silicon tandem solar cells. *Nat Rev Mater* 8, 261–281 (2023).
17. Asghar, M. I., Zhang, J., Wang, H. & Lund, P. D. Device stability of perovskite solar cells – A review. *Renewable and Sustainable Energy Reviews* 77, 131–146 (2017).
18. Dixit, H. *et al.* Assessment of lead-free tin halide perovskite solar cells using J–V hysteresis. *Physica status solidi (a)* 219, 2100823 (2022).

19. Nassiri Nazif, K. *et al.* High-specific-power flexible transition metal dichalcogenide solar cells. *Nat Commun* 12, 7034 (2021).
20. Nazif, K. N. *Transition Metal Dichalcogenides for Next-Generation Photovoltaics*. (Stanford University, 2021).
21. Nassiri Nazif, K., Nitta, F. U., Daus, A., Saraswat, K. C. & Pop, E. Efficiency limit of transition metal dichalcogenide solar cells. *Commun Phys* 6, 367 (2023).
22. Tiedje, T. O. M., Yablonovitch, E. L. I., Cody, G. D. & Brooks, B. G. Limiting efficiency of silicon solar cells. *IEEE Trans Electron Devices* 31, 711–716 (1984).
23. Freunek, M., Freunek, M. & Reindl, L. M. Maximum efficiencies of indoor photovoltaic devices. *IEEE J Photovolt* 3, 59–64 (2013).
24. Handbook, I. E. S. L. IES Lighting Handbook. *New York: Illuminating Engineering Society* (1966).
25. Michael, P. R., Johnston, D. E. & Moreno, W. A conversion guide: solar irradiance and lux illuminance. *Journal of Measurements in Engineering* 8, 153–166 (2020).
26. Went, C. M. *et al.* A new metal transfer process for van der Waals contacts to vertical Schottky-junction transition metal dichalcogenide photovoltaics. *Sci Adv* 5, eaax6061 (2019).
27. Green, M. A. Solar cell fill factors: General graph and empirical expressions. *Solid State Electronics* 24, 788–789 (1981).
28. Talla, V. *et al.* Lora backscatter: Enabling the vision of ubiquitous connectivity. *Proc ACM Interact Mob Wearable Ubiquitous Technol* 1, 1–24 (2017).
29. Dementyev, A., Hodges, S., Taylor, S. & Smith, J. Power consumption analysis of Bluetooth Low Energy, ZigBee and ANT sensor nodes in a cyclic sleep scenario. in *2013 IEEE International Wireless Symposium (IWS)* 1–4 (Beijing, China, 2013).
30. Singh, R. *et al.* Danger in the Dark: Stability of Perovskite Solar Cells with Varied Stoichiometries and Morphologies Stressed at Various Conditions. *ACS Appl Mater Interfaces* 16, 27450–27462 (2024).
31. Fojt, M., Teo, W. Z. & Pumera, M. Environmental impact and potential health risks of 2D nanomaterials. *Environ Sci Nano* 4, 1617–1633 (2017).
32. Das, S. *et al.* Transistors based on two-dimensional materials for future integrated circuits. *Nat Electron* 4, 786–799 (2021).
33. Dorow, C. J. *et al.* Gate length scaling beyond Si: mono-layer 2D channel FETs robust to short channel effects. in *2022 International Electron Devices Meeting (IEDM)* 5–7 (San Francisco, CA, 2022).
34. Chung, Y.-Y. *et al.* First Demonstration of GAA Monolayer-MoS2 Nanosheet nFET with 410 μA/μm ID 1 V VD at 40 nm gate length. in *2022 International Electron Devices Meeting (IEDM)* 34–35 (San Francisco, CA, 2022).
35. Wu, X. *et al.* Dual gate synthetic MoS2 MOSFETs with 4.56 μF/cm2 channel capacitance, 320 μS/μm Gm and 420 μA/μm Id at 1V Vd/100nm Lg. in *2021 IEEE International Electron Devices Meeting (IEDM)* 4–7 (San Francisco, CA, 2021).
36. Neilson, K. M. *et al.* Toward Mass Production of Transition Metal Dichalcogenide Solar Cells: Scalable Growth of Photovoltaic-Grade Multilayer WSe2 by Tungsten Selenization. *ACS Nano* 18, 24819–24828 (2024).

# Supplementary Information

## Transition Metal Dichalcogenide Solar Cells for Indoor Energy Harvesting

Frederick U. Nitta,[1,2] Koosha Nassiri Nazif,[1] and Eric Pop,[1,2,3*]

[1]Dept. of Electrical Engineering, Stanford University, Stanford, CA 94305, USA

[2]Dept. of Materials Science and Engineering, Stanford University, Stanford, CA 94305, USA

[3]Dept. of Applied Physics, Stanford University, Stanford, CA 94305, USA

*Corresponding author email: epop@stanford.edu

This supplementary information includes:

- **Supplementary Note 1.** Extended detailed balance method incorporating radiative, Auger, SRH recombination, and free carrier absorption, and different indoor spectra.
- **Supplementary Note 2.** Extrapolation of halogen lamp spectrum using blackbody radiation formula.
- **Supplementary Figure 1.** Extrapolation of halogen spectrum at 500 lux.
- **Supplementary Table 1.** Input power density across different lighting conditions.
- **Supplementary Figures 2 – 5.** Short-circuit current density ($J_{SC}$), open-circuit voltage ($V_{OC}$), fill factor (FF), and output power ($P_{out}$) of thin-film TMD solar cells under CFL illumination.
- **Supplementary Figures 6 – 9.** $J_{SC}$, $V_{OC}$, FF, and $P_{out}$ of thin-film TMD solar cells under LED illumination.
- **Supplementary Figures 10 – 13.** $J_{SC}$, $V_{OC}$, FF, and $P_{out}$ of thin-film TMD solar cells under halogen illumination.
- **Supplementary Figures 14 – 17.** $J_{SC}$, $V_{OC}$, FF, and $P_{out}$ of thin-film TMD solar cells under AM 1.5 G illumination.
- **Supplementary Table 2.** Literature reports on indoor photovoltaic devices.

**Supplementary Note 1. Extended detailed balance method incorporating radiative, Auger, SRH recombination, and free carrier absorption, and different indoor spectra.**

Based on the Tiedje-Yablonovitch model[37], the detailed balance equation that defines the current density–voltage ($J-V$) characteristics of an optimized solar cell with an intrinsic or lightly-doped absorber film, characterized by equal electron ($N$) and hole density ($P$) under illumination, is applicable under the presence of radiative emission, Auger recombination, and free carrier absorption. The equation that governs this $J-V$ relationship is as follows:

$$\left(\alpha_1 + \frac{1}{4n^2L}\right)\exp\left(\frac{eV}{kT}\right)\int\int a_2(E)b_n(E,T)dE\,d\Omega + CN^3 = \frac{J_{\text{in}}}{eL}(1-f) \quad (1)$$

Here, $\alpha_1$ is the free carrier absorption coefficient, $n$ is the refractive index of the absorber film, $L$ is the thickness of the film, $e$ is the elementary charge, $V$ is the output voltage, $k$ is the Boltzmann constant, $T$ is temperature, $a_2(E)$ is the absorptance (absorption probability) at photon energy $E$, $b_\text{n}(E,T)dEd\Omega$ is the flux of black-body photons for a photon energy interval $dE$ and solid angle $d\Omega$ in a medium with refractive index of $n$, $C$ is the Auger coefficient, $N$ is the electron (and hole) density, $\frac{J_{\text{in}}}{eL}$ is the volume rate of generation of electron-hole pairs by the sun, and $f$ is the fraction of the incident solar flux that is drawn off as current into the external circuit. $a_2(E)$, $b_\text{n}(E,T)$ and $J_{\text{in}}$ are defined as:

$$a_2(E) = \frac{\alpha_2(E)}{\alpha_2(E) + \alpha_1(E) + \frac{1}{4n^2L}} \quad (2)$$

$$b_\text{n}(E,T) = \frac{2}{h^3}\frac{n^2}{c^2}E^2\exp\left(\frac{1}{\frac{E}{kT}-1}\right) \quad (3)$$

$$J_{\text{in}} = \int eS(E)a_2(E)dE \quad (4)$$

where $\alpha_2(E)$ is the optical absorption coefficient at photon energy $E$, $h$ is the Planck constant, $c$ is the speed of light in vacuum, and $S(E)$ is the one of the four indoor spectra we employed for this work. In **Equation (1)**, the terms on the left-hand side sequentially represent the rates of free carrier absorption, radiative emission, and Auger recombination. Conversely, the terms on the right-hand side describe the generation rate of electron-hole pairs and the solar cell's output current, respectively. To include Shockley-Read-Hall (SRH) recombination into this model, we add SRH recombination rate $U_{\text{SRH}}$ to the left-hand side of **Equation (1)**:

$$\left(\alpha_1 + \frac{1}{4n^2L}\right)\exp\left(\frac{eV}{kT}\right)\int\int a_2(E)b_n(E,T)dE\,d\Omega + CN^3 + U_{\text{SRH}} = \frac{J_{\text{in}}}{eL}(1-f) \quad (5)$$

Carrier lifetimes associated with each recombination mechanism, $\tau_{\mathrm{e}}$ and $\tau_{\mathrm{h}}$ for electrons and holes, respectively, can be defined as:

$$\tau_{\mathrm{e}} = \frac{\Delta N}{U} \tag{6}$$

$$\tau_{h} = \frac{\Delta P}{U} \tag{7}$$

where $\Delta N$ and $\Delta P$ are the disturbances of the electron and hole populations from their equilibrium values $N_0$ and $P_0$, respectively, and $U$ is the recombination rate. For an intrinsic or lightly-doped absorber film under illumination:

$$N = P \gg N_0, P_0 \tag{8}$$

$$\Delta N = \Delta P \approx N \tag{9}$$

Thus, the SRH recombination rate can be written as:

$$U_{\mathrm{SRH}} = \frac{N}{\tau_{\mathrm{SRH}}} \tag{10}$$

Combining **Equations (5)** and **(10)** leads to the following:

$$\left(\alpha_1 + \frac{1}{4n^2L}\right)\exp\left(\frac{eV}{kT}\right)\int\int a_2(E)b_n(E,T)dE\,d\Omega + CN^3 + \frac{N}{\tau_{\mathrm{SRH}}} = \frac{J_{\mathrm{in}}}{eL}(1-f) \tag{11}$$

**Equation (11)** represents the detailed balance equation governing the current density–voltage characteristics of optimized solar cell having an intrinsic or lightly-doped absorber film (i.e., $N = P$ under illumination) in the presence of radiative emission, Auger recombination, free carrier absorption, and SRH recombination with the characteristic carrier lifetime $\tau_{\mathrm{SRH}}$. In the absence of free carrier absorption, **Equation (11)** simplifies further:

$$J_0\exp\left(\frac{eV}{kT}\right) + eLCN_i^3 exp\left(\frac{3eV}{2kT}\right) + \frac{eL}{\tau_{\mathrm{SRH}}}N_i\exp\left(\frac{eV}{2kT}\right) = J_{\mathrm{in}}(1-f) \tag{12}$$

where $N_i$ is the intrinsic carrier density and $J_0$ is defined as:

$$J_0 = e\pi\int b_1(E)a_2(E)dE \tag{13}$$

To derive the current density–voltage characteristics of the solar cell, $f$ is varied from zero to one, which corresponds to output current density ($J$) of zero to $J_{\mathrm{in}}$. The output voltage ($V$) is then calculated by solving **Equation (12).** From these $J-V$ characteristics, performance metrics are extracted as follows:

$$V_{\mathrm{OC}} = V(J=0) \tag{14}$$

$$J_{\mathrm{SC}} = J(V = 0) \tag{15}$$

$$P_{\mathrm{MPP}} = \max(I \cdot V) = I \cdot V(\frac{d(I \cdot V)}{dV} = 0) \tag{16}$$

$$FF = \frac{P_{\mathrm{MPP}}}{V_{\mathrm{OC}} \cdot J_{\mathrm{SC}}} \tag{17}$$

$$PCE = \frac{P_{MPP}}{P_{in}} \tag{18}$$

where $V_{\mathrm{OC}}$ is the open-circuit voltage, $J_{\mathrm{SC}}$ is the short-circuit current density, $P_{\mathrm{MPP}}$ is power density at maximum power point (MPP), $FF$ is the fill factor, and $PCE$ is power conversion efficiency of the solar cell. $P_{in}$, the input power density, depends on the spectra and the lux level, and the calculated values are listed in **Supplementary Table 1**.

**Supplementary Note 2. Extrapolation of halogen lamp spectrum using blackbody radiation formula.**

Halogen lamps, which operate by heating a tungsten filament to high temperatures within a halogen gas, closely mimic the emission characteristics of a blackbody radiator[38]. This similarity allows for the use of the blackbody radiation formula to extend the lamp's spectral data[39], which did not cover the full range of photon energies emitted, especially at lower energy (higher wavelengths). We employed Planck's law of blackbody radiation[40], expressed as

$$B_\nu(\nu,T) = \frac{2h\nu^3}{c^2}\frac{1}{e^{\frac{h\nu}{kT}}-1} \tag{19}$$

where $h$ is Planck's constant, $\nu$ is the frequency of radiation, $c$ is the speed of light in vacuum, $k$ is Boltzmann's constant. The absolute temperature $T$ of the blackbody is 2847 K, which was determined from the fitting of experimental data to **Equation 19**. This fit is shown in **Supplementary Figure 1**.

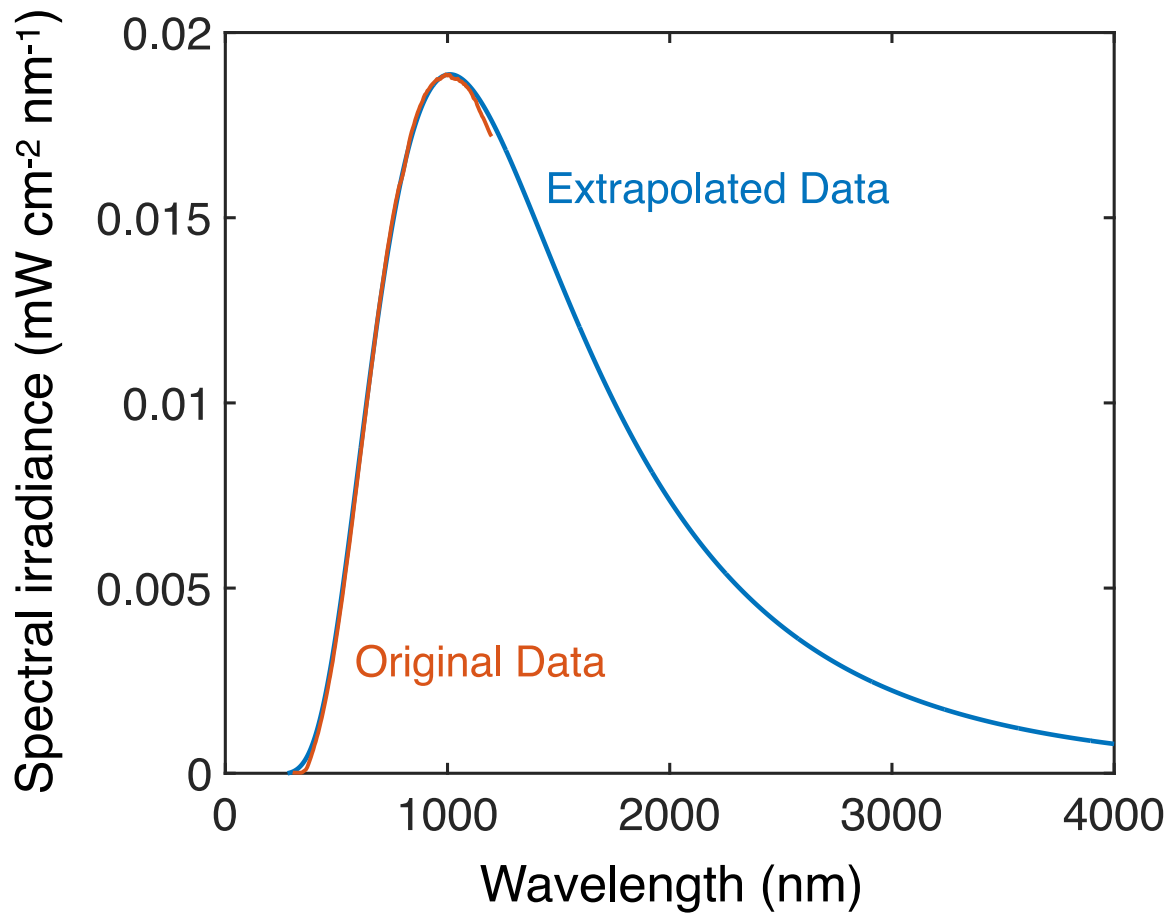

**Supplementary Figure 1 | Extrapolation of halogen spectrum** at 500 lux. The original (experimental) data is in orange, our extrapolation in blue. The extrapolation was performed using the blackbody radiation formula detailed in **Supplementary Note 2**. The precision of this method was confirmed by an R-square value of 1.00 and root mean square error (RMSE) of $2.64 \times 10^{-5}$.

**Supplementary Table 1 | Input power density across different lighting conditions.** The input power densities (mW $cm^{-2}$) for different light sources across varying illumination levels.

| | Parking garage (50 lux) | Warehouse (150 lux) | Office (400 lux) | Retail store (500 lux) |
|---|---|---|---|---|
| Compact fluorescent lamp (CFL) | 0.145 | 0.434 | 1.157 | 1.447 |
| Light-emitting diode (LED) | 0.175 | 0.526 | 1.402 | 1.752 |
| Halogen | 2.605 | 7.814 | 20.837 | 26.047 |
| AM 1.5 G | 0.432 | 1.297 | 3.459 | 4.324 |

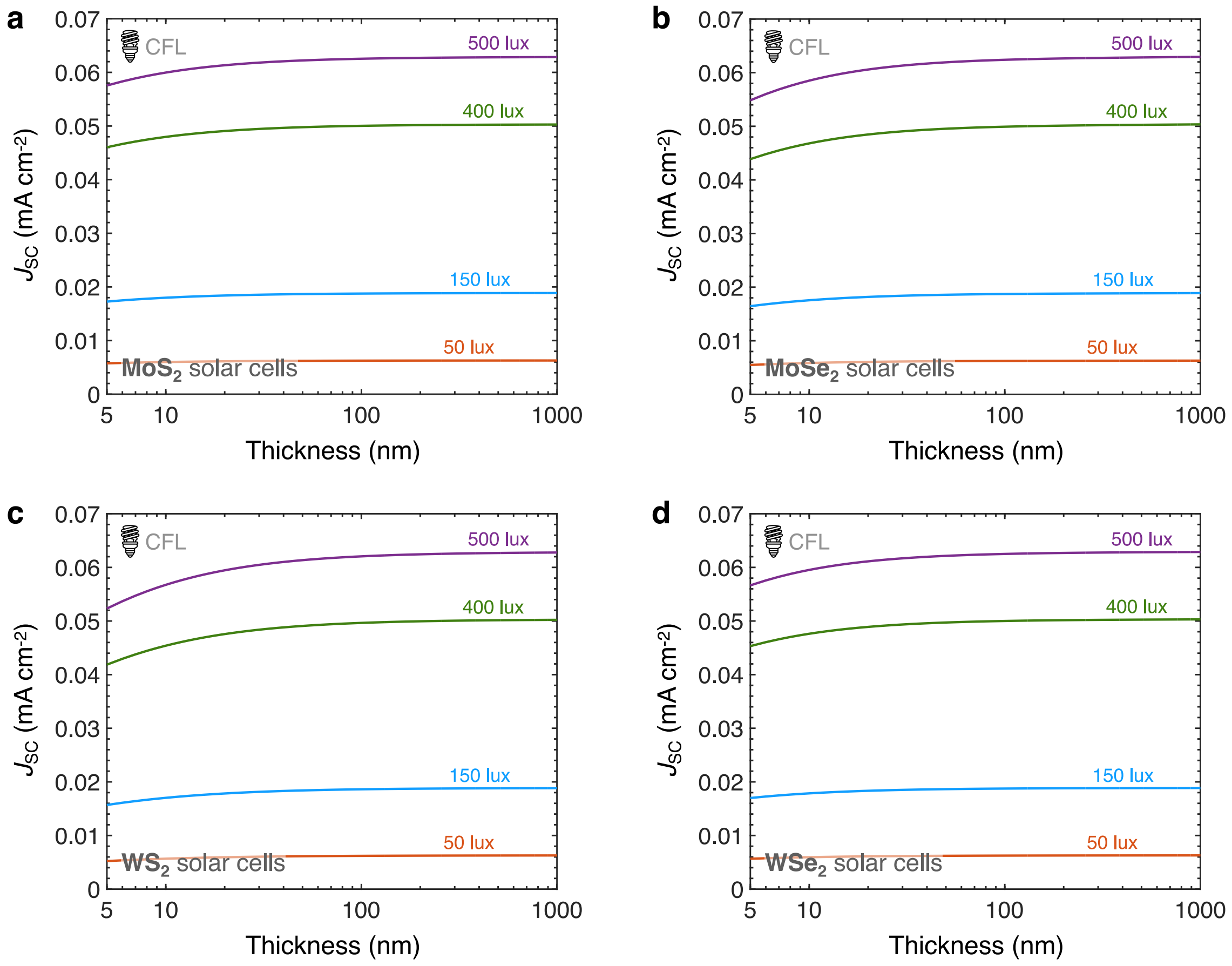

**Supplementary Figure 2 | Short-circuit current density ($J_{SC}$) of thin-film TMD solar cells under compact fluorescent lamp (CFL) illumination**. $J_{SC}$ of **a**, $MoS_2$, **b**, $MoSe_2$, **c**, $WS_2$, and **d**, $WSe_2$ solar cells as a function of the TMD (absorber) film thickness and CFL illumination intensity at 300 K. Four CFL illumination intensities correspond to the four colors, as labeled (e.g. purple is at 500 lux). The $J_{SC}$ exhibits minimal improvement with increased film thickness at lower light intensities such as parking garages (50 lux) or warehouses (150 lux), reflecting near-unity absorption for higher-energy photons (above ~2.0 eV)[21,41,42], which dominate the CFL spectrum at these intensities. In contrast, at higher intensities found in office (400 lux) or retail store (500 lux) settings, $J_{SC}$ shows more improvement as the film thickness increases, attributable to enhanced absorption of lower-energy photons (below ~2.0 eV). At higher film thicknesses, the $J_{SC}$ values for all materials start to plateau as the absorption reaches near-complete across the CFL spectrum's most relevant energy range, from ~1.5 eV to ~3.5 eV. This indicates that the materials have achieved their maximum potential for light absorption in this spectrum, and making the films thicker beyond this point will not increase the photocurrent.

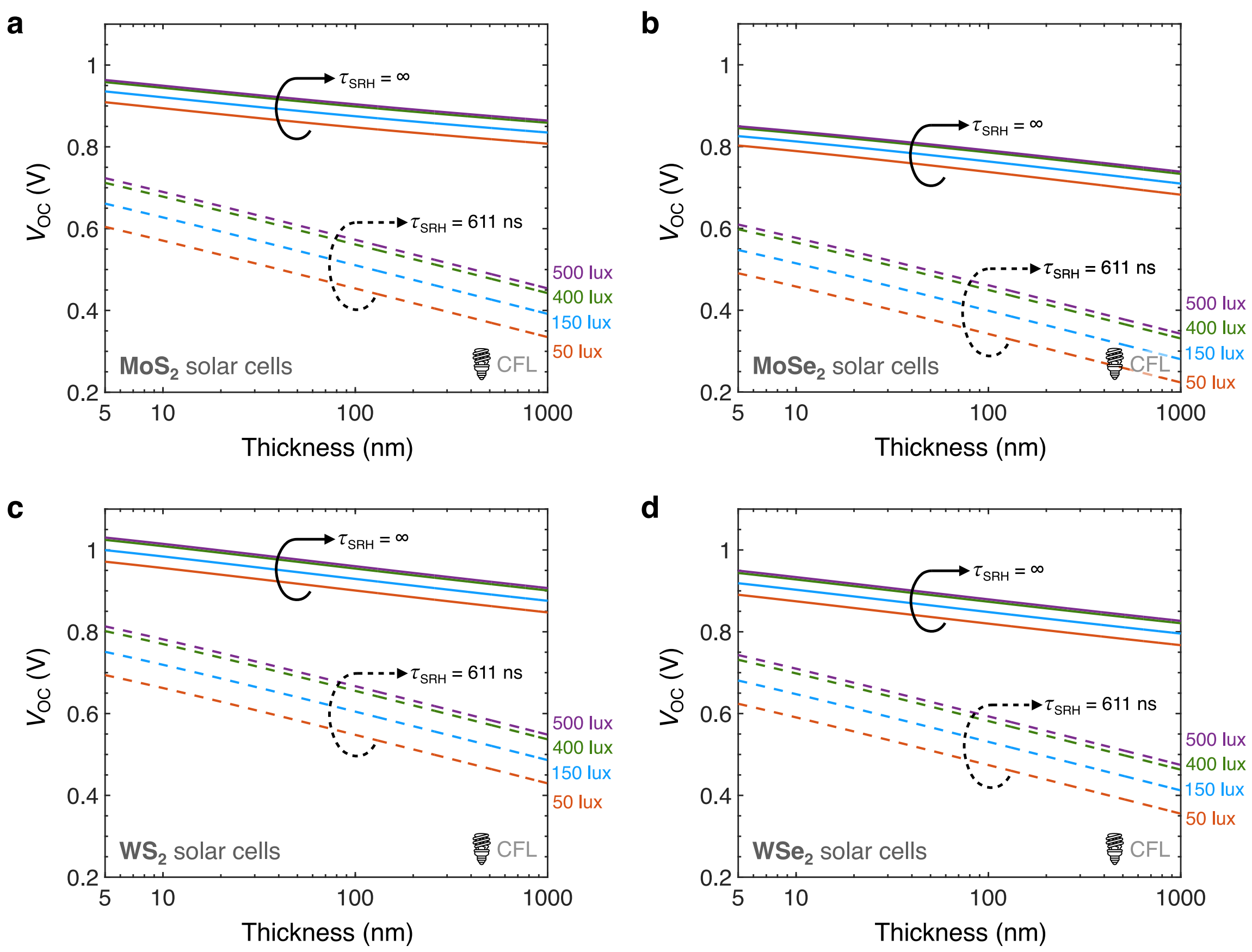

**Supplementary Figure 3 | Open-circuit voltage ($V_{OC}$) of thin-film TMD solar cells under CFL illumination.** $V_{OC}$ of **a,** $MoS_2$, **b,** $MoSe_2$, **c,** $WS_2$, and **d,** $WSe_2$ solar cells as a function of TMD film thickness, material quality ($\tau_{SRH}$), and CFL illumination intensity at 300 K. $\tau_{SRH}$, Shockley-Read-Hall (SRH) lifetime. Solid lines are in the limit defect-free material (no SRH recombination), dashed lines with $\tau_{SRH}$ = 611 ns. Four CFL illumination intensities correspond to the four colors, as labeled (e.g. purple dashed and solid are at 500 lux). $V_{OC}$ exhibits a logarithmic decline with decreasing light intensity, reflecting its direct proportionality to the logarithm of the photocurrent, which is dependent on light intensity. Additionally, $V_{OC}$ decreases with increasing film thickness, more notably for films with an SRH lifetime of 611 ns. This decrease can be partially attributed to a shift in the absorption threshold towards lower photon energies as films become thicker[21], effectively reducing the band gap and thus $V_{OC}$. The increase in film thickness also leads to higher light absorption and charge carrier densities, escalating the likelihood of both radiative and non-radiative recombination events. The pronounced $V_{OC}$ decline with $\tau_{SRH}$ = 611 ns underscores the impact of material quality on maintaining high $V_{OC}$ across various lighting conditions.

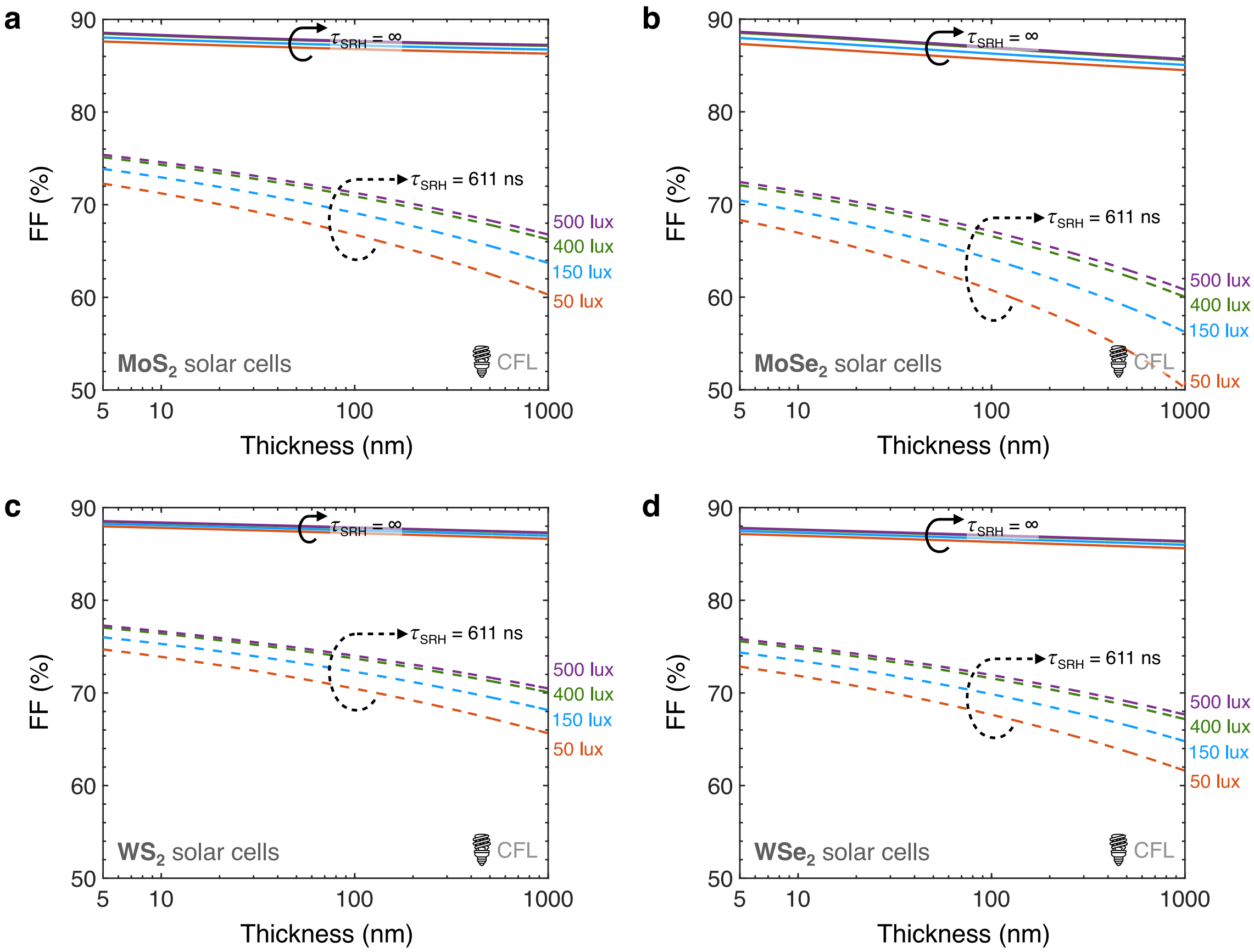

**Supplementary Figure 4 | Fill factor (FF) of thin-film TMD solar cells under CFL illumination.** FF of **a,** $MoS_2$, **b,** $MoSe_2$, **c,** $WS_2$, and **d,** $WSe_2$ solar cells as a function of TMD film thickness, material quality ($\tau_{SRH}$), and CFL illumination intensity at 300 K. $\tau_{SRH}$, Shockley-Read-Hall (SRH) lifetime. Solid lines are in the limit defect-free material (no SRH recombination), dashed lines with $\tau_{SRH}$ = 611 ns. Four CFL illumination intensities correspond to the four colors, as labeled (e.g. purple dashed and solid are at 500 lux). It is established that a higher $V_{OC}$ generally leads to a higher FF due to reduced recombination losses[27]. When considering a $\tau_{SRH}$ of 611 ns, the drop in both $V_{OC}$ and FF with increased film thickness is more pronounced, particularly under lower light intensities. This is because lower light intensities reduce carrier generation, and although SRH recombination itself decreases with lower carrier densities, the relative impact of each recombination event is greater when fewer carriers are available. As a result, both $V_{OC}$ and FF decline more sharply in these conditions, especially in thicker films where defects are more prominent. The more pronounced the SRH recombination, the greater the impact on the ideality factor of the diode and consequently on FF. This lines up with the understanding that the closer the ideality factor is to unity (as in the case of infinite $\tau_{SRH}$), the higher the FF, whereas dominant SRH recombination leads to an ideality factor of 2, reducing the FF. The dependence of FF on $V_{OC}$ also explains why FF decreases with increasing film thickness. As shown in **Supplementary Figure 3**, $V_{OC}$ decreases with increasing film thickness, and FF follows the same trend.

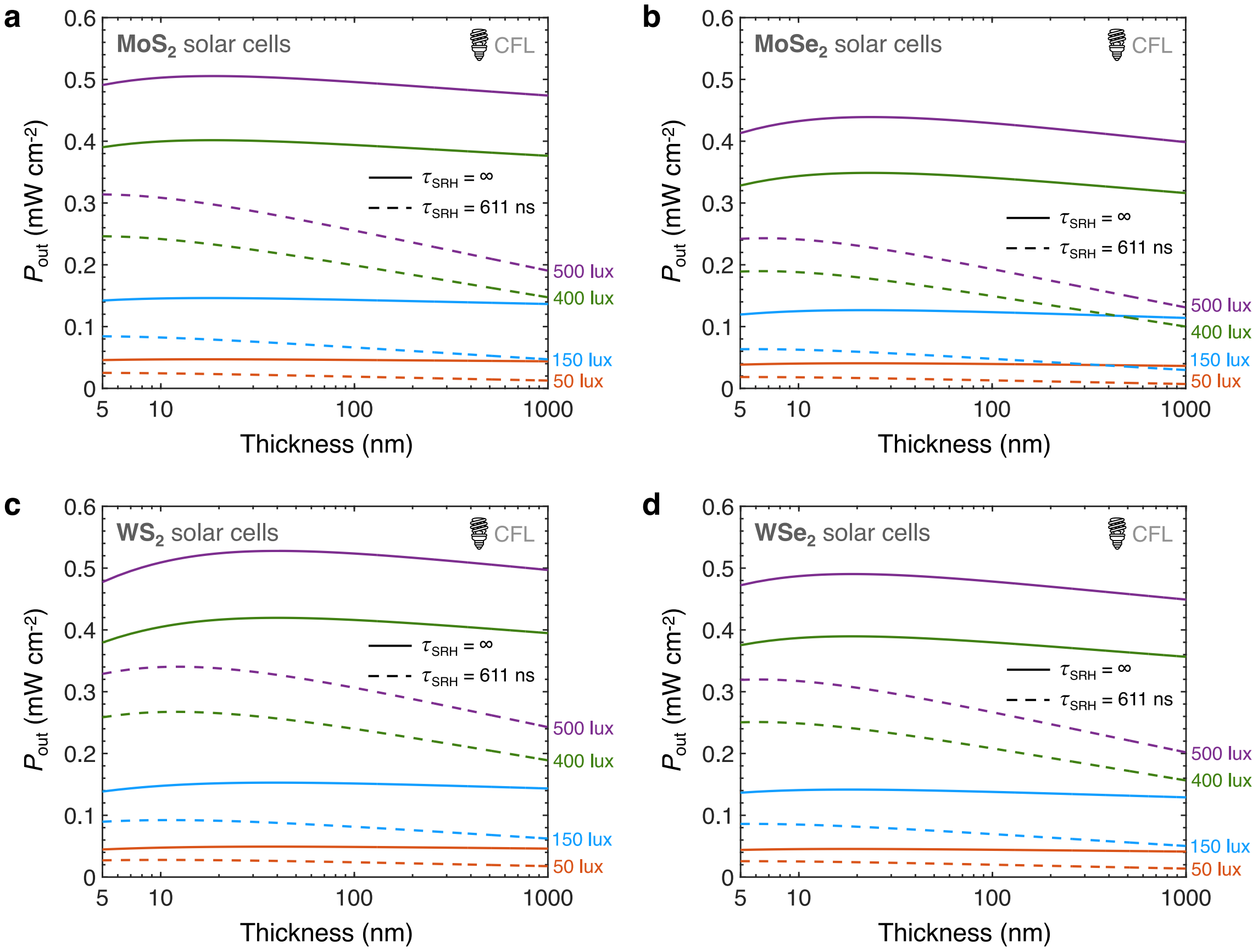

**Supplementary Figure 5 | Output power ($P_{out}$) of thin-film TMD solar cells under CFL illumination.** $P_{out}$ of **a,** $MoS_2$, **b,** $MoSe_2$, **c,** $WS_2$, and **d,** $WSe_2$ solar cells as a function of TMD film thickness, material quality ($\tau_{SRH}$), and CFL illumination intensity at 300 K. $\tau_{SRH}$, Shockley-Read-Hall (SRH) lifetime. Solid lines are in the limit defect-free material (no SRH recombination), dashed lines with $\tau_{SRH}$ = 611 ns. Four CFL illumination intensities correspond to the four colors, as labeled (e.g. purple dashed and solid are at 500 lux). Since $P_{out}$ of solar cells is the product of $J_{SC}$, $V_{OC}$, and FF, these trends are explained by the $J_{SC}$, $V_{OC}$, and fill factor trends in **Supplementary Figure 2, Supplementary Figure 3**, and **Supplementary Figure 4**, respectively. As observed, the output power curves exhibit an inverted U-shape, which can be explained by the competing in-fluences of $J_{SC}$ (**Supplementary Figure 2**) and the product of $V_{OC}$ (**Supplementary Figure 3**) and FF (**Supplementary Figure 4**). As the film thickness increases, $J_{SC}$ improves due to enhanced light absorption; however, this benefit is counterbalanced by the degradation of $V_{OC}$ and FF, which is particularly pronounced when $\tau_{SRH}$ is finite at 611 ns. At lower light intensities (e.g., 50 lux), $J_{SC}$ does not increase substantially with increasing thickness (**Supplementary Figure 2**). Similarly, for the infinite $\tau_{SRH}$ scenario (no SRH recombination), $V_{OC}$ and FF experience a mild drop with thickness (**Supplementary Figure 3** and **Supplementary Figure 4**). Therefore, the improvement in $J_{SC}$ with thickness is offset by losses in $V_{OC}$ and FF, resulting in a constant $P_{out}$ across the range of film thicknesses considered. SRH recombination, however, leads to steeper decline in $V_{OC}$ and fill factor with thickness (**Supplementary Figure 3** and **Supplementary Figure 4**), leading to decreasing $P_{out}$ with increasing thickness. This behavior underscores the critical role that material quality and defect minimization plays in optimizing the power output of TMD-based solar cells.

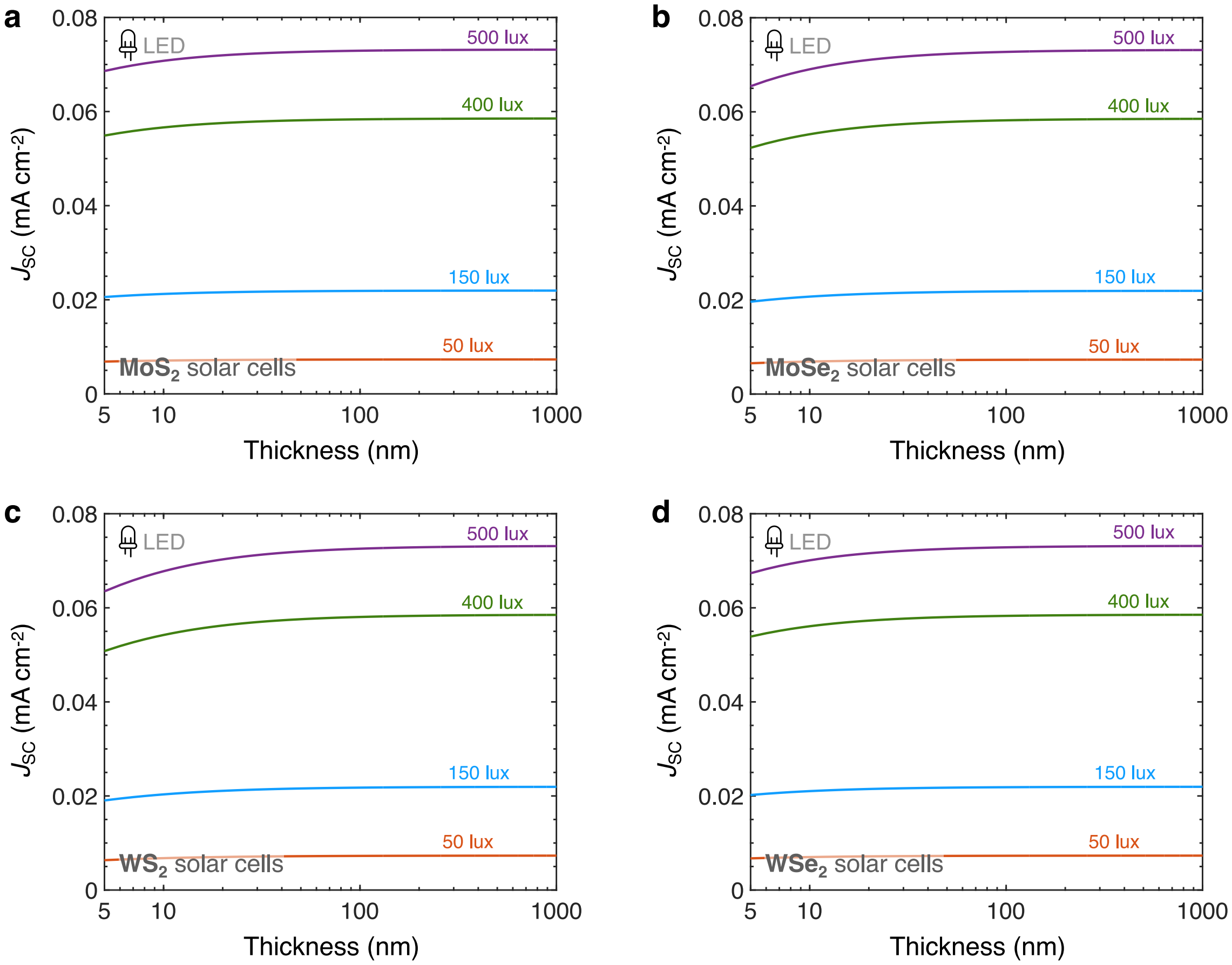

**Supplementary Figure 6 | Short-circuit current density ($J_{SC}$) of thin-film TMD solar cells under LED illumination**. $J_{SC}$ of **a,** $MoS_2$, **b,** $MoSe_2$, **c,** $WS_2$, and **d,** $WSe_2$ solar cells as a function of the TMD (absorber) film thickness and LED illumination intensity at 300 K. Four CFL illumination intensities correspond to the four colors, as labeled (e.g. purple is at 500 lux). Similar to CFL lighting, under LED illumination, $J_{SC}$ enhancement with film thickness is modest at low intensities, like in parking garages (50 lux) and warehouses (150 lux), but better at higher intensities found in offices (400 lux) and retail stores (500 lux). The slightly higher $J_{SC}$ values under LED compared to CFL lighting can be attributed to the LED spectrum aligning less with the CIE Photopic Luminosity Function than CFL. Consequently, LED sources require more emitted light to achieve the same lux levels (**Supplementary Table 1**), which increases the available photon flux for energy conversion. For all TMD materials, $J_{SC}$ varies more distinctly at lower thicknesses under LED light. The LED spectrum's broader photon energy range at higher intensities allows for increased $J_{SC}$ in thicker films due to absorption of lower-energy photons. At higher thicknesses, $J_{SC}$ plateaus for all materials, indicating maximum potential absorption in the relevant energy range of LED light.

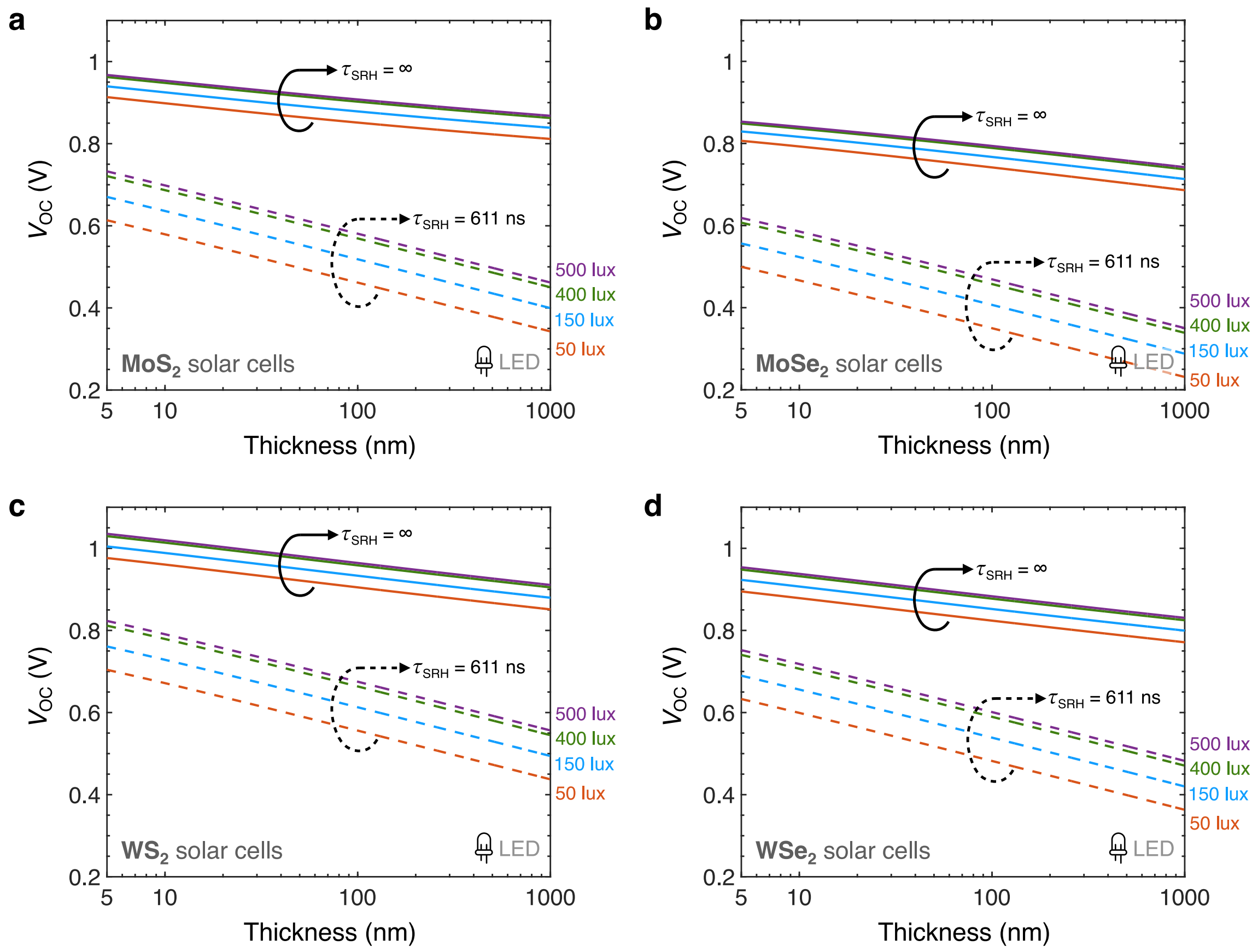

**Supplementary Figure 7 | Open-circuit voltage ($V_{OC}$) of thin-film TMD solar cells under LED illumination.** $V_{OC}$ of **a,** $MoS_2$, **b,** $MoSe_2$, **c,** $WS_2$, and **d,** $WSe_2$ solar cells as a function of TMD film thickness, material quality ($\tau_{SRH}$), and LED illumination intensity at 300 K. $\tau_{SRH}$, Shockley-Read-Hall (SRH) lifetime. Solid lines are in the limit defect-free material (no SRH recombination), dashed lines with $\tau_{SRH}$ = 611 ns. Four CFL illumination intensities correspond to the four colors, as labeled (e.g. purple dashed and solid are at 500 lux). With increasing TMD film thickness, there is a noticeable reduction in $V_{OC}$, particularly when considering a finite SRH lifetime. This reduction is consistent with what was observed under CFL lighting. As film thickness grows, the absorption threshold shifts, diminishing the effective band gap and thus the $V_{OC}$[21]. The resulting higher carrier densities from this shift elevate recombination rates, with non-radiative processes having a more significant impact under finite SRH conditions. This relationship showcases the critical effect of material defects on $V_{OC}$, underlining the importance of advancing material quality for improved TMD solar cell performance.

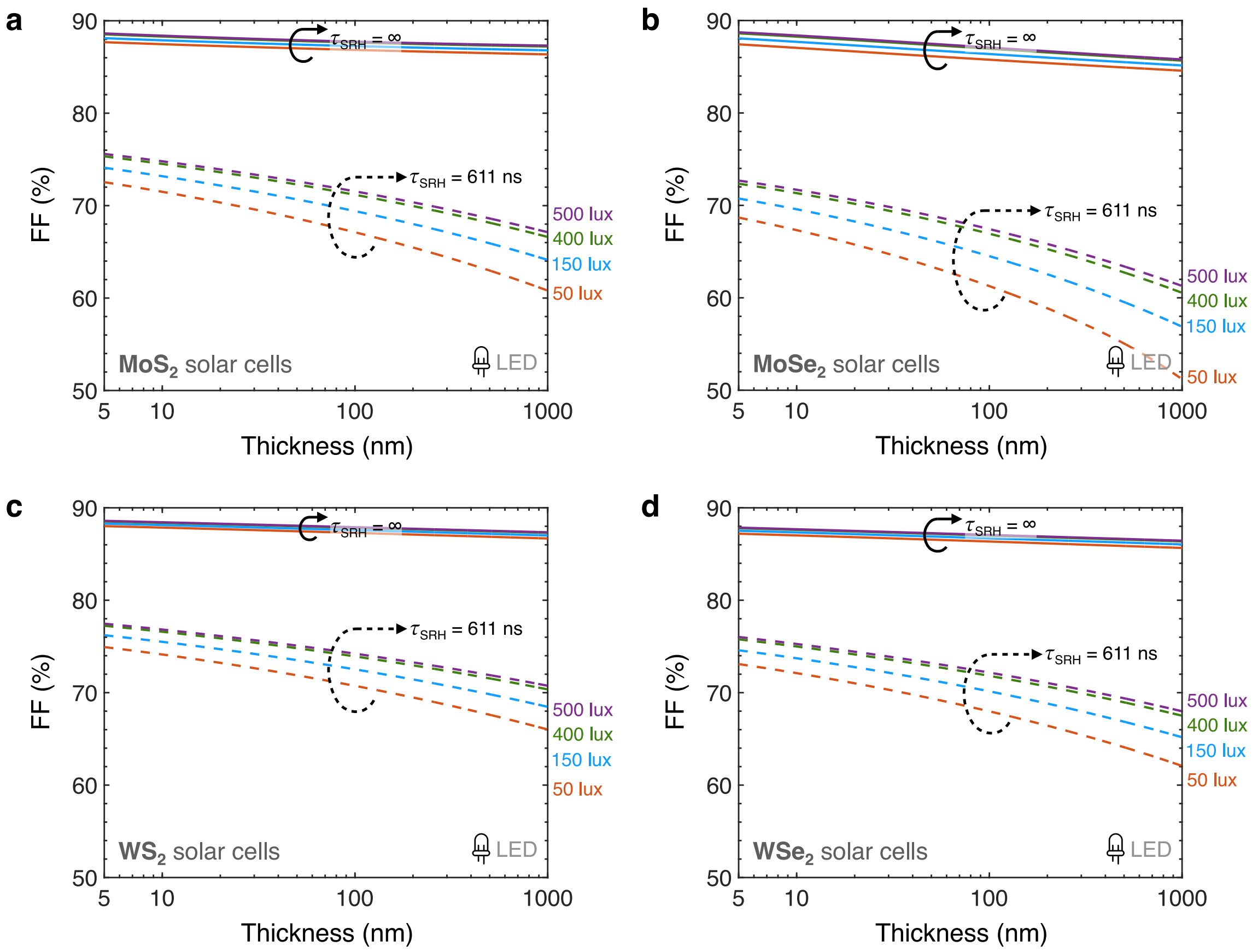

**Supplementary Figure 8 | Fill factor (FF) of thin-film TMD solar cells under LED illumination.** FF of **a,** $MoS_2$, **b,** $MoSe_2$, **c,** $WS_2$, and **d,** $WSe_2$ solar cells as a function of TMD film thickness, material quality ($\tau_{SRH}$), and LED illumination intensity at 300 K. $\tau_{SRH}$, Shockley-Read-Hall (SRH) lifetime. Solid lines are in the limit defect-free material (no SRH recombination), dashed lines with $\tau_{SRH}$ = 611 ns. As illumination intensity increases, $V_{OC}$ and consequently FF generally improve due to increased carrier generation, which helps reduce the relative impact of recombination losses. Under infinite $\tau_{SRH}$ conditions, FF is almost independent of light intensity and film thickness, indicating a dominant role for non-radiative recombination. The consistency of FF values across all thicknesses in the infinite $\tau_{SRH}$ case contrasts with the decline observed at finite $\tau_{SRH}$. When $\tau_{SRH}$ is finite (e.g., 611 ns), FF shows a more pronounced decrease with increasing film thickness, particularly at lower light intensities. This is because thicker films absorb more light, generating more carriers, but also increasing the likelihood of carrier recombination at defect sites, especially when material quality is lower (as indicated by a finite $\tau_{SRH}$). At lower light intensities, while fewer carriers are generated overall, the relative impact of defects is greater because fewer carriers are available to be collected.

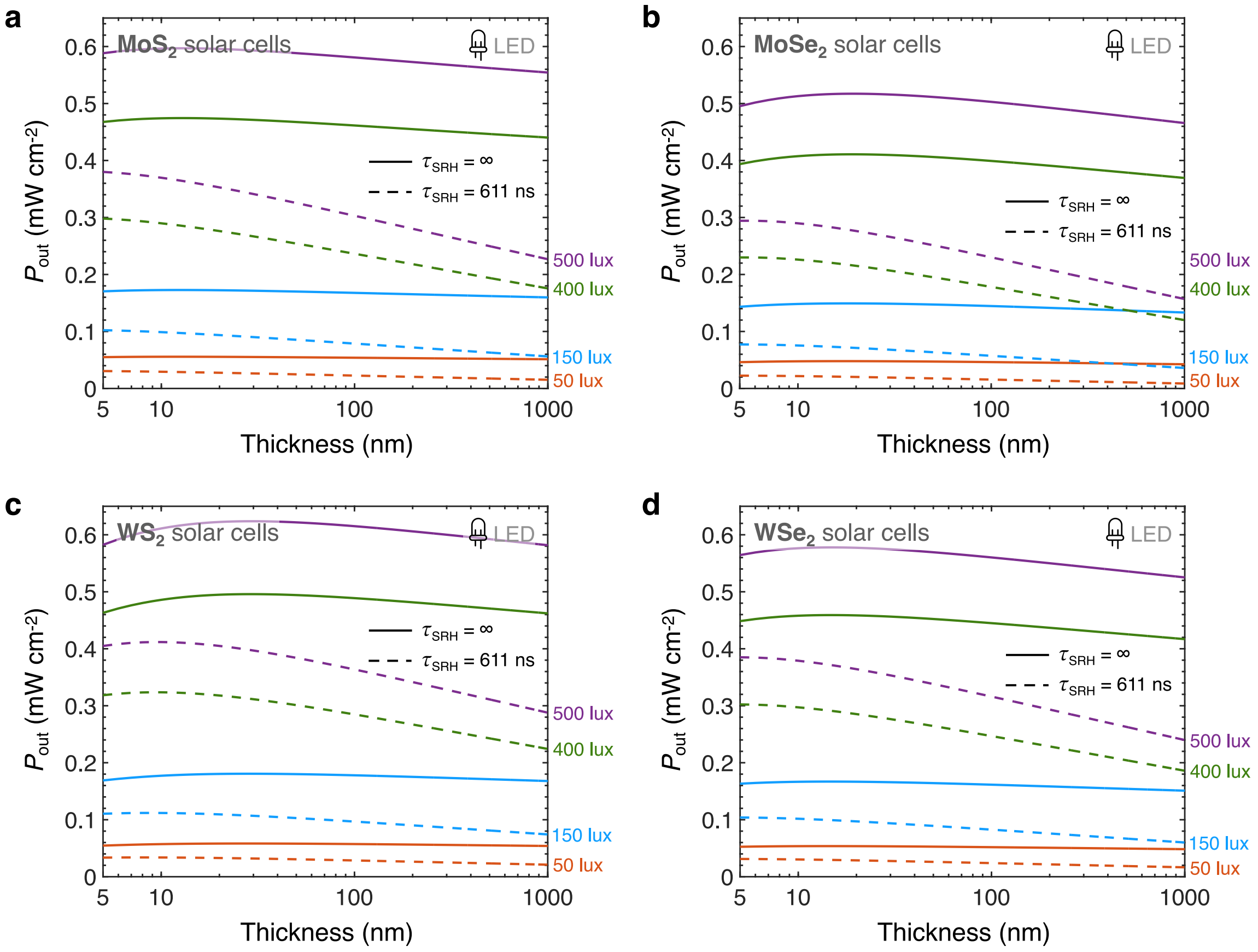

**Supplementary Figure 9 | Output power ($P_{out}$) of thin-film TMD solar cells under LED illumination.** $P_{out}$ of **a,** $MoS_2$, **b,** $MoSe_2$, **c,** $WS_2$, and **d,** $WSe_2$ solar cells as a function of TMD film thickness, material quality ($\tau_{SRH}$), and LED illumination intensity at 300 K. $\tau_{SRH}$, Shockley-Read-Hall (SRH) lifetime. Solid lines are in the limit defect-free material (no SRH recombination), dashed lines with $\tau_{SRH}$ = 611 ns. Four CFL illumination intensities correspond to the four colors, as labeled (e.g. purple dashed and solid are at 500 lux). Under LED lighting, similar to CFL illumination, $P_{out}$ demonstrates a dependence on $J_{SC}$, $V_{OC}$, and FF behaviors in **Supplementary Figure 6**, **Supplementary Figure 7**, and **Supplementary Figure 8**, respectively. At lower light intensities (e.g., 50 lux), the $P_{out}$'s peak is less distinct due to the modest increase in $J_{SC}$ with film thickness. $P_{out}$ is influenced by the trade-off between enhanced $J_{SC}$ from increased thickness and the reductions in $V_{OC}$ and FF, which is especially noticeable at finite $\tau_{SRH}$. For infinite $\tau_{SRH}$, $P_{out}$ changes less across film thicknesses since $V_{OC}$ and FF are more constant. This results in a flatter $P_{out}$ curve, indicating that defects and SRH recombination have a larger effect on $P_{out}$ than radiative and Auger recombination. The variation in $P_{out}$ between the different material qualities ($\tau_{SRH}$ values) underlines the importance of material quality and defect control.

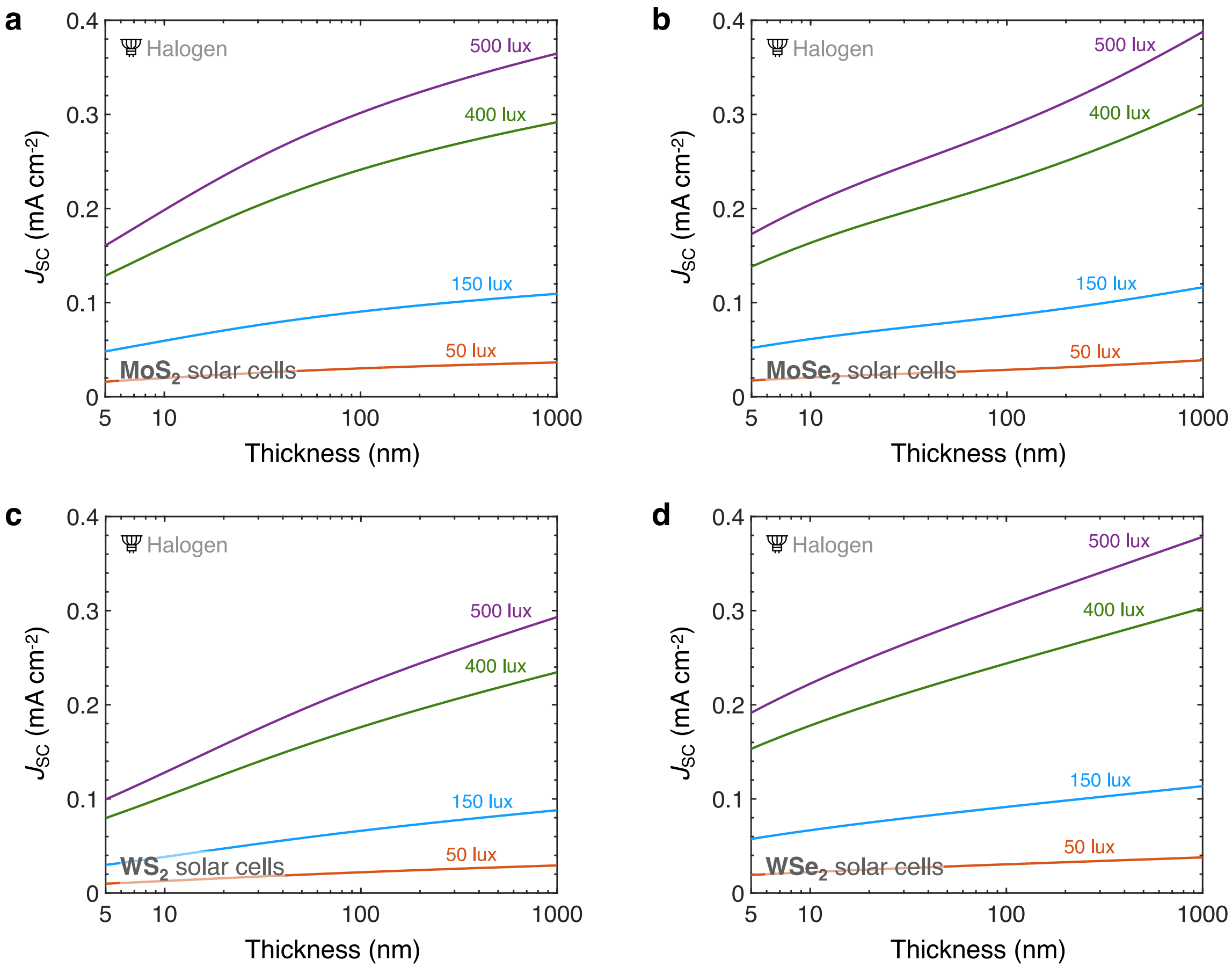

**Supplementary Figure 10 | Short-circuit current density ($J_{SC}$) of thin-film TMD solar cells under halogen illumination**. $J_{SC}$ of **a,** $MoS_2$, **b,** $MoSe_2$, **c,** $WS_2$, and **d,** $WSe_2$ solar cells as a function of the TMD (absorber) film thickness and halogen illumination intensity at 300 K. Four CFL illumination intensities correspond to the four colors, as labeled (e.g. purple is at 500 lux). Like with CFL lighting, there is less $J_{SC}$ improvement at lower light intensities under halogen lighting, yet the increase is more compared to CFL due to the halogen bulb's broader spectral coverage. Under halogen lighting, $J_{SC}$ is comparatively higher than under CFL due to the spectral characteristics of halogen light. Halogen bulbs emit across a broader spectral range with less overlap with the CIE Photopic Luminosity Function than CFL sources, requiring more intensity to achieve the same perceived brightness (**Supplementary Table 1**). This wider distribution of energy across the spectrum leads to enhanced $J_{SC}$ values for TMD solar cells.

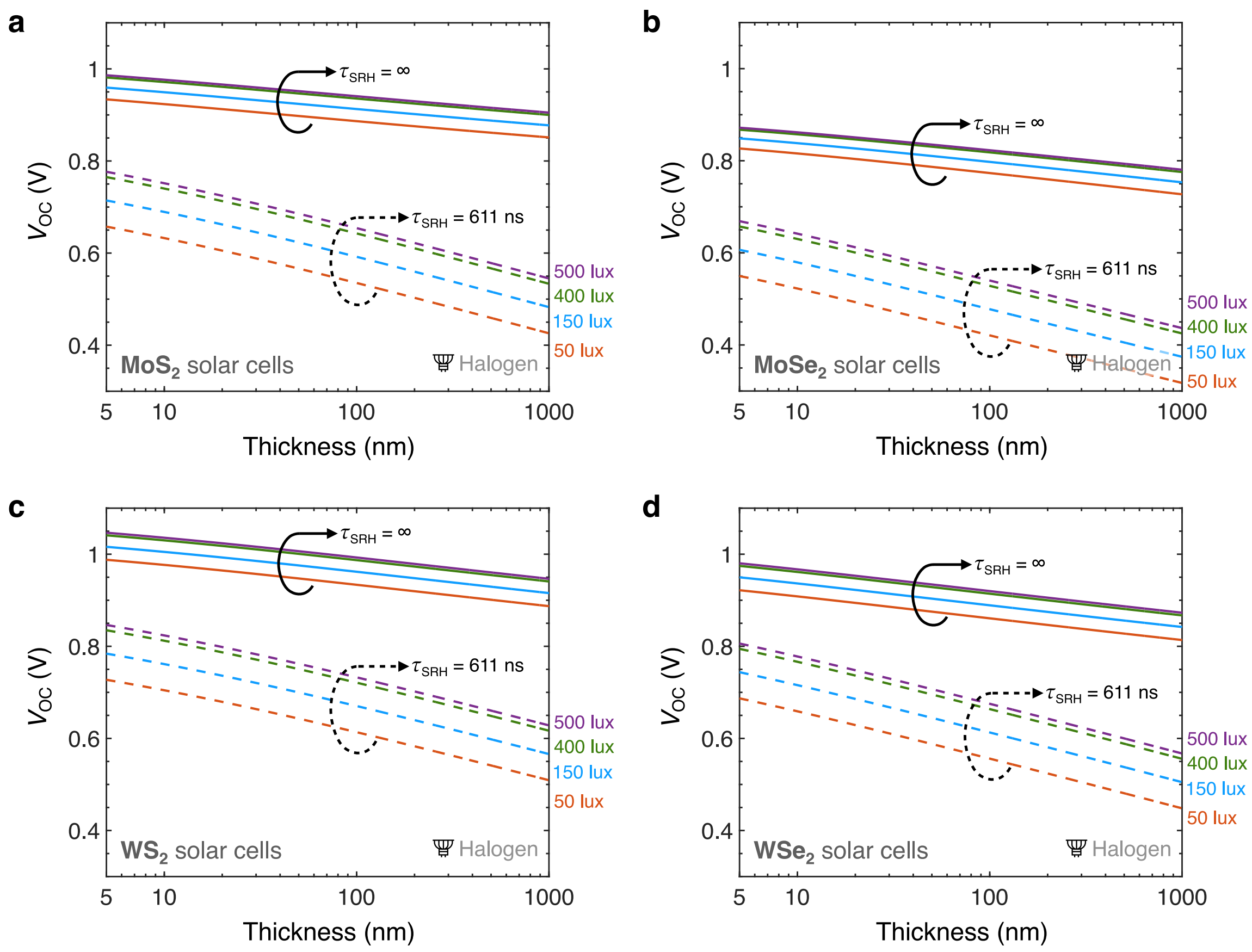

**Supplementary Figure 11 | Open-circuit voltage ($V_{OC}$) of thin-film TMD solar cells under halogen illumination.** $V_{OC}$ of **a,** $MoS_2$, **b,** $MoSe_2$, **c,** $WS_2$, and **d,** $WSe_2$ solar cells as a function of TMD film thickness, material quality ($\tau_{SRH}$), and halogen illumination intensity at 300 K. $\tau_{SRH}$, Shockley-Read-Hall (SRH) lifetime. Solid lines are in the limit defect-free material (no SRH recombination), dashed lines with $\tau_{SRH}$ = 611 ns. Four CFL illumination intensities correspond to the four colors, as labeled (e.g. purple dashed and solid are at 500 lux). Similar to CFL and LED lighting, $V_{OC}$ decreases with an increase in film thickness, more noticeably at the finite SRH lifetime of 611 ns due to increased recombination at defect sites. This reduction is attributed to the shift in absorption threshold with thicker films, resulting in a lower effective band gap and higher carrier densities that enhance recombination[21].

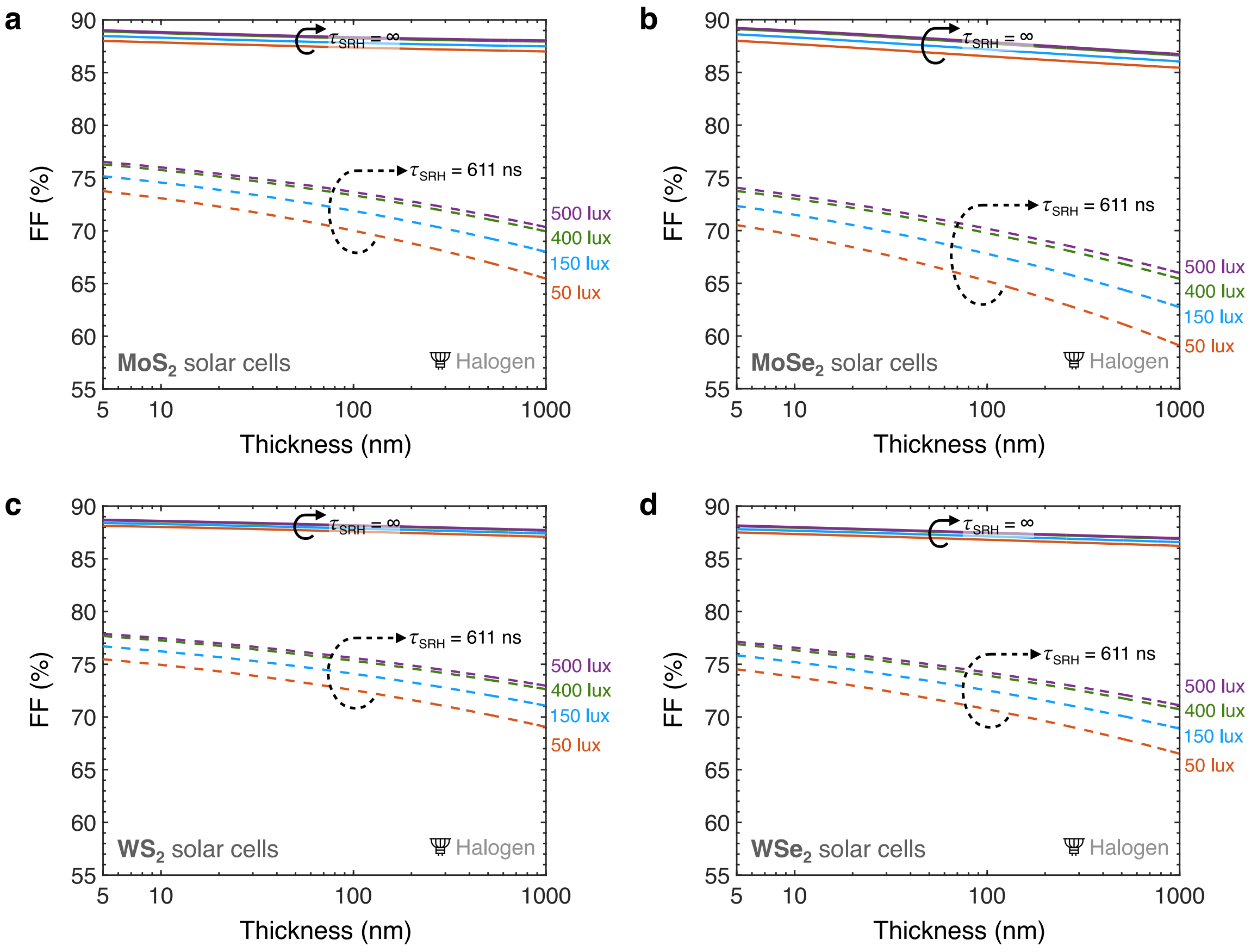

**Supplementary Figure 12 | Fill factor (FF) of thin-film TMD solar cells under halogen illumination.** FF of **a,** $MoS_2$, **b,** $MoSe_2$, **c,** $WS_2$, and **d,** $WSe_2$ cells as a function of TMD film thickness, material quality ($\tau_{SRH}$), and halogen illumination intensity at 300 K. $\tau_{SRH}$, Shockley-Read-Hall (SRH) lifetime. Solid lines are in the limit defect-free material (no SRH recombination), dashed lines with $\tau_{SRH}$ = 611 ns. Four CFL illumination intensities correspond to the four colors, as labeled (e.g. purple dashed and solid are at 500 lux). With a $\tau_{SRH}$ of 611 ns, the decrease in $V_{OC}$ and consequently FF becomes more pronounced as film thickness increases, particularly under lower light intensities. This is because, at lower light intensities, fewer carriers are generated, and any losses due to recombination at defect sites have a disproportionately larger impact on the FF. In thicker films, where there are more potential defect sites, this effect is even more pronounced. Across the TMD materials, FF is quite uniform at all thicknesses for an infinite $\tau_{SRH}$, suggesting minimal radiative and Auger recombination. In contrast, for a finite $\tau_{SRH}$, FF begin the same at lower thicknesses but spread out as the films get thicker. This spread results from how each material's carrier density affects non-radiative recombination, which alters the FF.

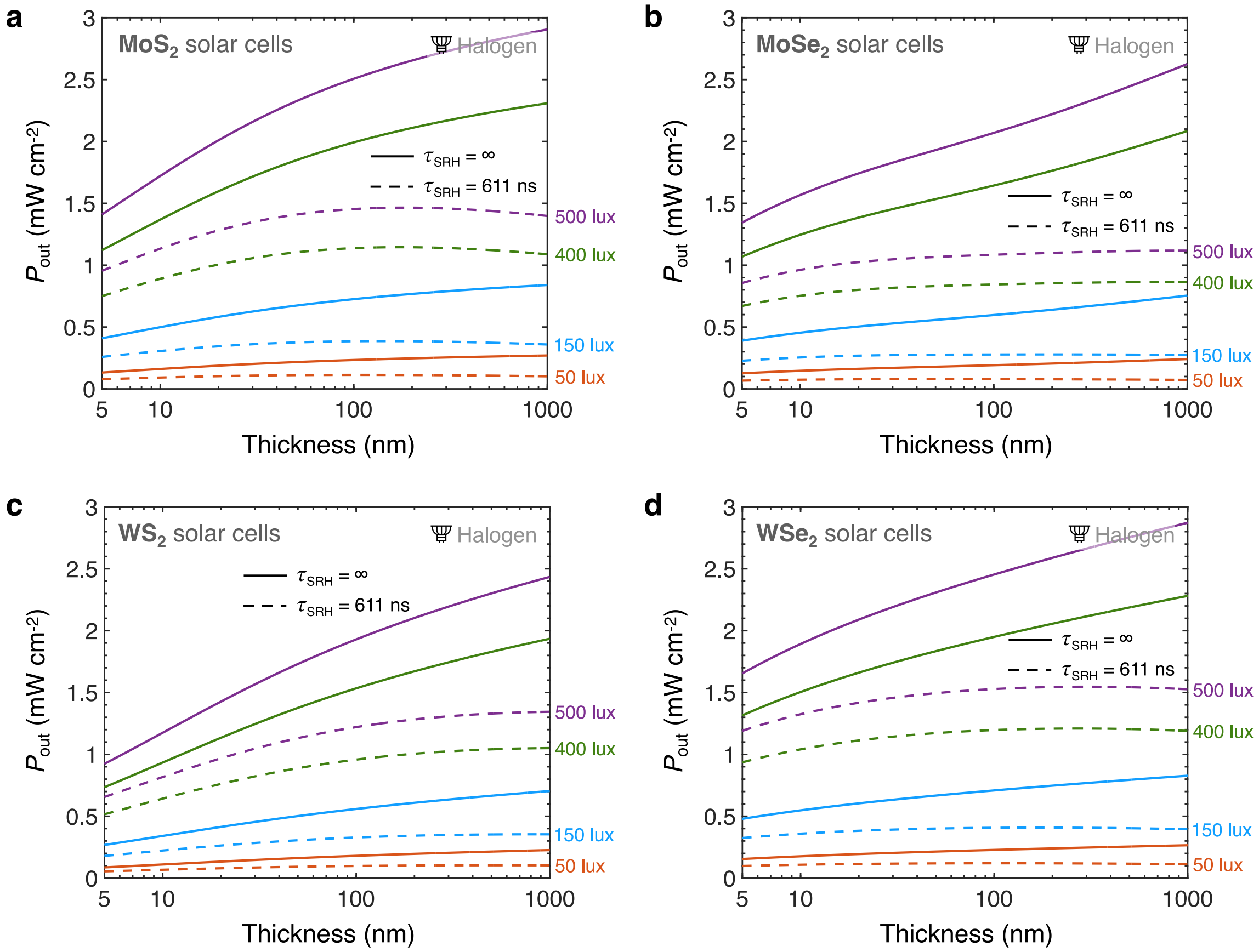

**Supplementary Figure 13 | Output power ($P_{out}$) of thin-film TMD solar cells under halogen illumination.** $P_{out}$ of **a,** $MoS_2$, **b,** $MoSe_2$, **c,** $WS_2$, and **d,** $WSe_2$ solar cells as a function of TMD film thickness, material quality ($\tau_{SRH}$), and halogen illumination intensity at 300 K. $\tau_{SRH}$, Shockley-Read-Hall (SRH) lifetime. Solid lines are in the limit defect-free material (no SRH recombination), dashed lines with $\tau_{SRH}$ = 611 ns. Four CFL illumination intensities correspond to the four colors, as labeled (e.g. purple dashed and solid are at 500 lux). For an infinite SRH lifetime, $P_{out}$ for all materials consistently rises without a peak, implying that the positive effects of increased $J_{SC}$ with thickness (**Supplementary Figure 10**) outweigh the negative impacts on $V_{OC}$ (**Supplementary Figure 11**) and FF (**Supplementary Figure 12**). This trend suggests that halogen light, with its broad spectrum, may be effectively utilized by thicker TMD films without the penalties of increased recombination from defects. In contrast, for a finite $\tau_{SRH}$ of 611 ns, $MoS_2$ and $WSe_2$ display peaks in $P_{out}$ at certain thicknesses. This indicates that there are optimal thicknesses at which the benefits of increased absorption (and hence $J_{SC}$) surpass the detriments caused by higher recombination rates affecting $V_{OC}$ and FF, and beyond a certain film thickness, the additional material does not proportionally contribute to power generation under halogen illumination.

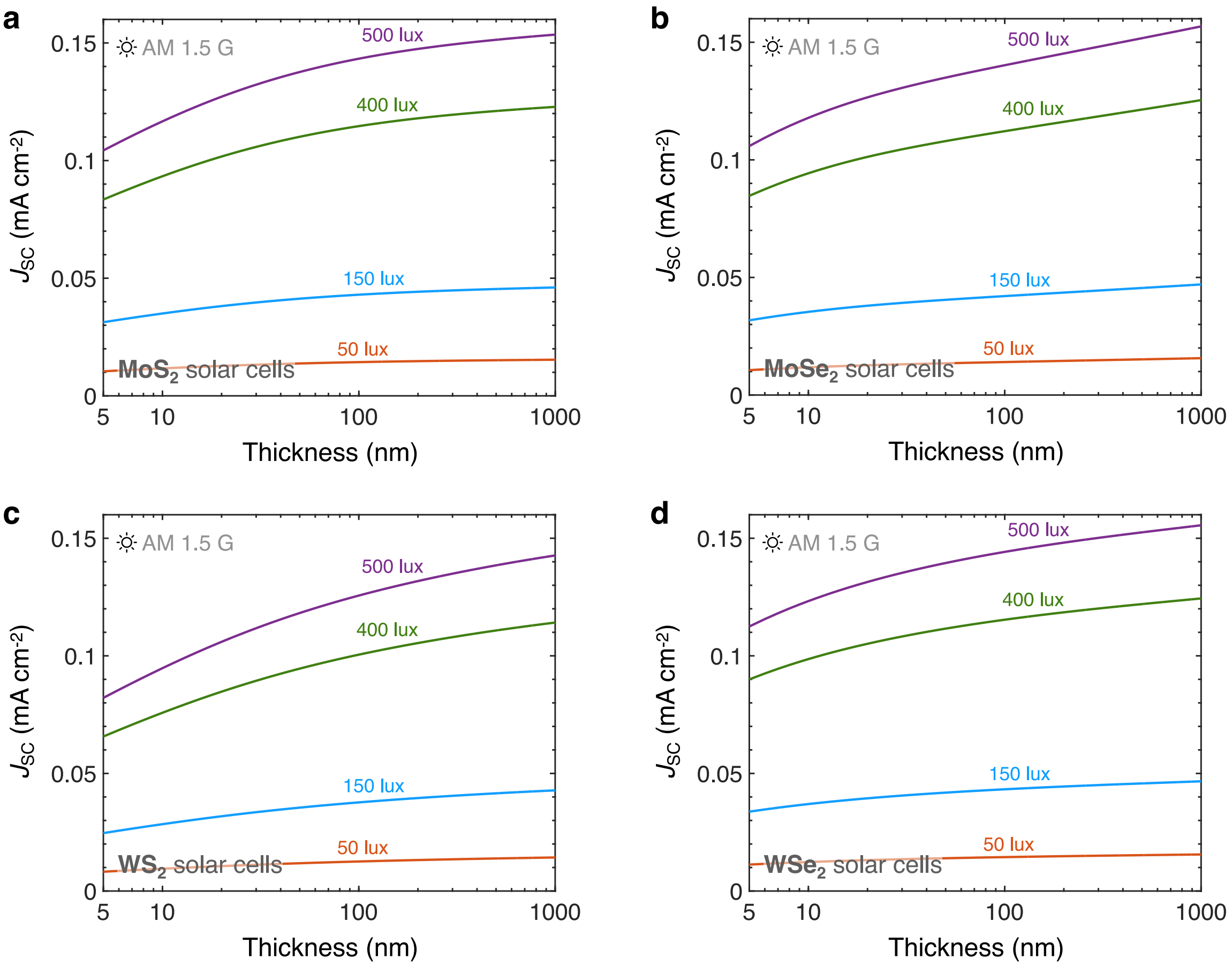

**Supplementary Figure 14 | Short-circuit current density ($J_{SC}$) of thin-film TMD solar cells under AM 1.5 G illumination**. $J_{SC}$ of **a,** $MoS_2$, **b,** $MoSe_2$, **c,** $WS_2$, and **d,** $WSe_2$ solar cells as a function of the TMD (absorber) film thickness and AM 1.5 G illumination intensity at 300 K. Four CFL illumination intensities correspond to the four colors, as labeled (e.g. purple is at 500 lux). The $J_{SC}$ values under AM 1.5 G are lower than those under halogen lighting because the AM 1.5 G spectrum has a greater overlap with the CIE Photopic Luminosity Function, resulting in less photon flux for the same lux levels. This characteristic necessitates a lower AM 1.5 G illumination intensity to match the defined indoor lighting scenarios (**Supplementary Table 1**), thus yielding a reduced $J_{SC}$.

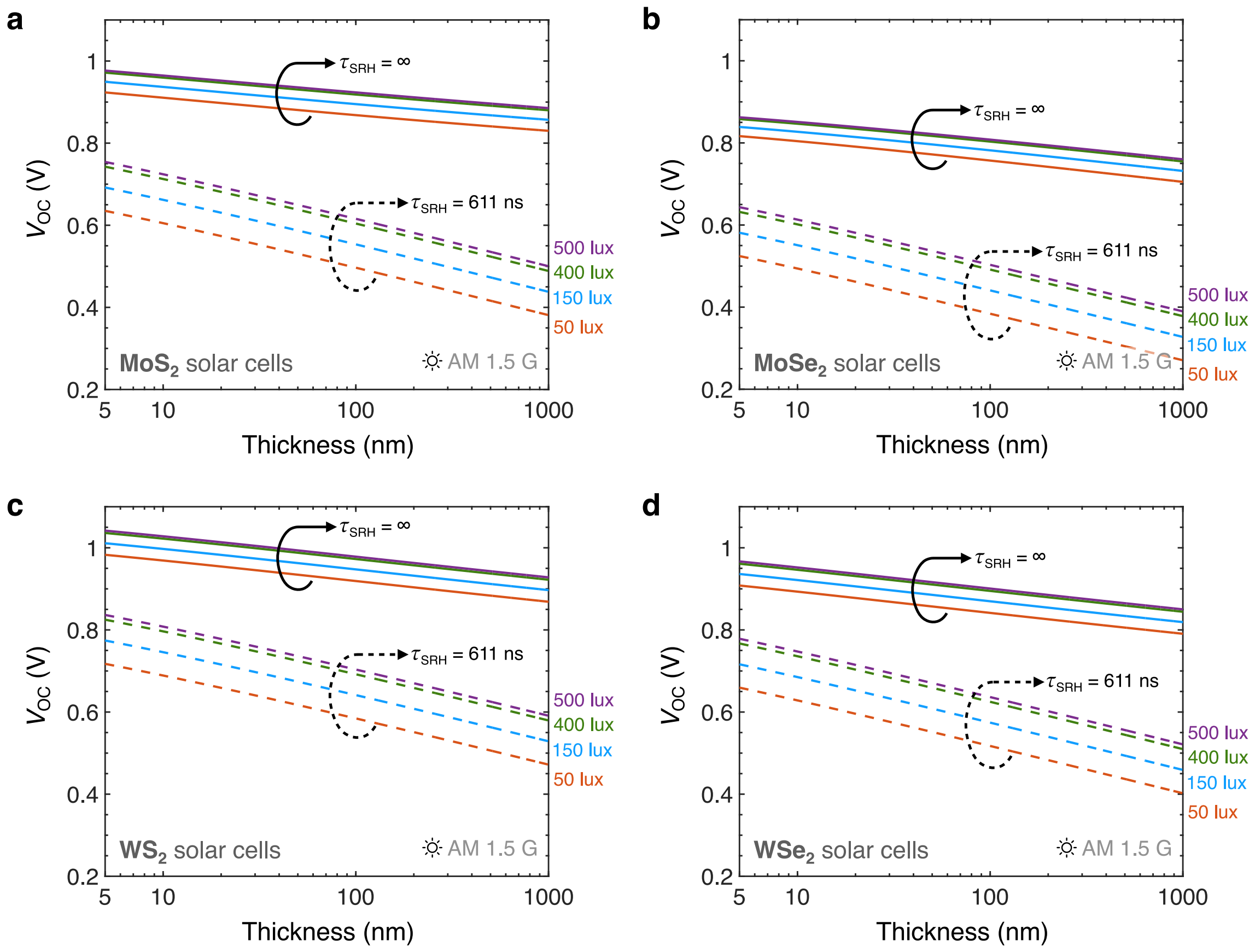

**Supplementary Figure 15 | Open-circuit voltage ($V_{OC}$) of thin-film TMD solar cells under AM 1.5 G illumination.** $V_{OC}$ of **a,** $MoS_2$, **b,** $MoSe_2$, **c,** $WS_2$, and **d,** $WSe_2$ solar cells as a function of TMD film thickness, material quality ($\tau_{SRH}$), and AM 1.5 G illumination intensity at 300 K. $\tau_{SRH}$, Shockley-Read-Hall (SRH) lifetime. Solid lines are in the limit defect-free material (no SRH recombination), dashed lines with $\tau_{SRH}$ = 611 ns. Four CFL illumination intensities correspond to the four colors, as labeled (e.g. purple dashed and solid are at 500 lux). With increasing film thickness, a noticeable decrease in $V_{OC}$ is observed, particularly when $\tau_{SRH}$ is set at 611 ns. This decline is attributed to a shift in the absorption threshold to lower photon energies as the TMD films thicken. As a result, the effective band gap decreases, which, along with the heightened absorption of thicker films, leads to increased carrier density and a greater chance of non-radiative recombination events, thereby reducing $V_{OC}$[21]. The trends align with those under other illumination conditions such as CFL, LED, and halogen, where the reduction in $V_{OC}$ is more significant with the presence of material defects as represented by the finite $\tau_{SRH}$.

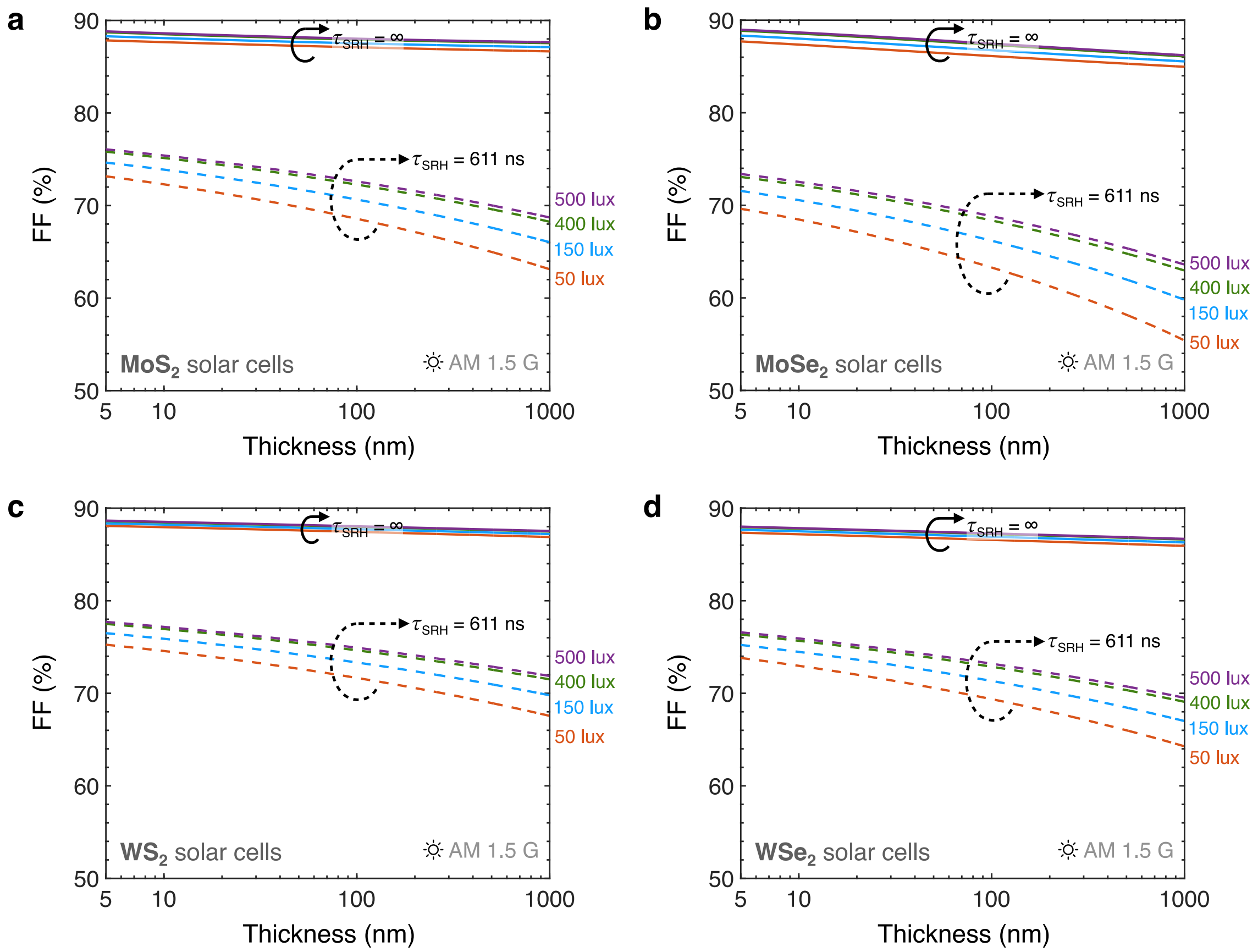

**Supplementary Figure 16 | Fill factor (FF) of thin-film TMD solar cells under AM 1.5 G illumination.** FF of **a,** $MoS_2$, **b,** $MoSe_2$, **c,** $WS_2$, and **d,** $WSe_2$ solar cells as a function of TMD film thickness, material quality ($\tau_{SRH}$), and AM 1.5 G illumination intensity at 300 K. $\tau_{SRH}$, Shockley-Read-Hall (SRH) lifetime. Solid lines are in the limit defect-free material (no SRH recombination), dashed lines with $\tau_{SRH}$ = 611 ns. Four CFL illumination intensities correspond to the four colors, as labeled (e.g. purple dashed and solid are at 500 lux). A finite $\tau_{SRH}$ of 611 ns leads to a marked decrease in $V_{OC}$ and thus FF with increasing film thickness. This trend largely mirrors behavior under CFL, halogen, and LED lighting, and it is consistent with the understanding that thicker films, while capturing more light, also enable more recombination events at defect sites. The FF reduction is more noticeable at higher thickness levels where defects have a greater influence on recombination. In contrast, FF remains relatively stable across various thicknesses when an infinite $\tau_{SRH}$ is assumed, indicating an ideal case with negligible defect recombination.

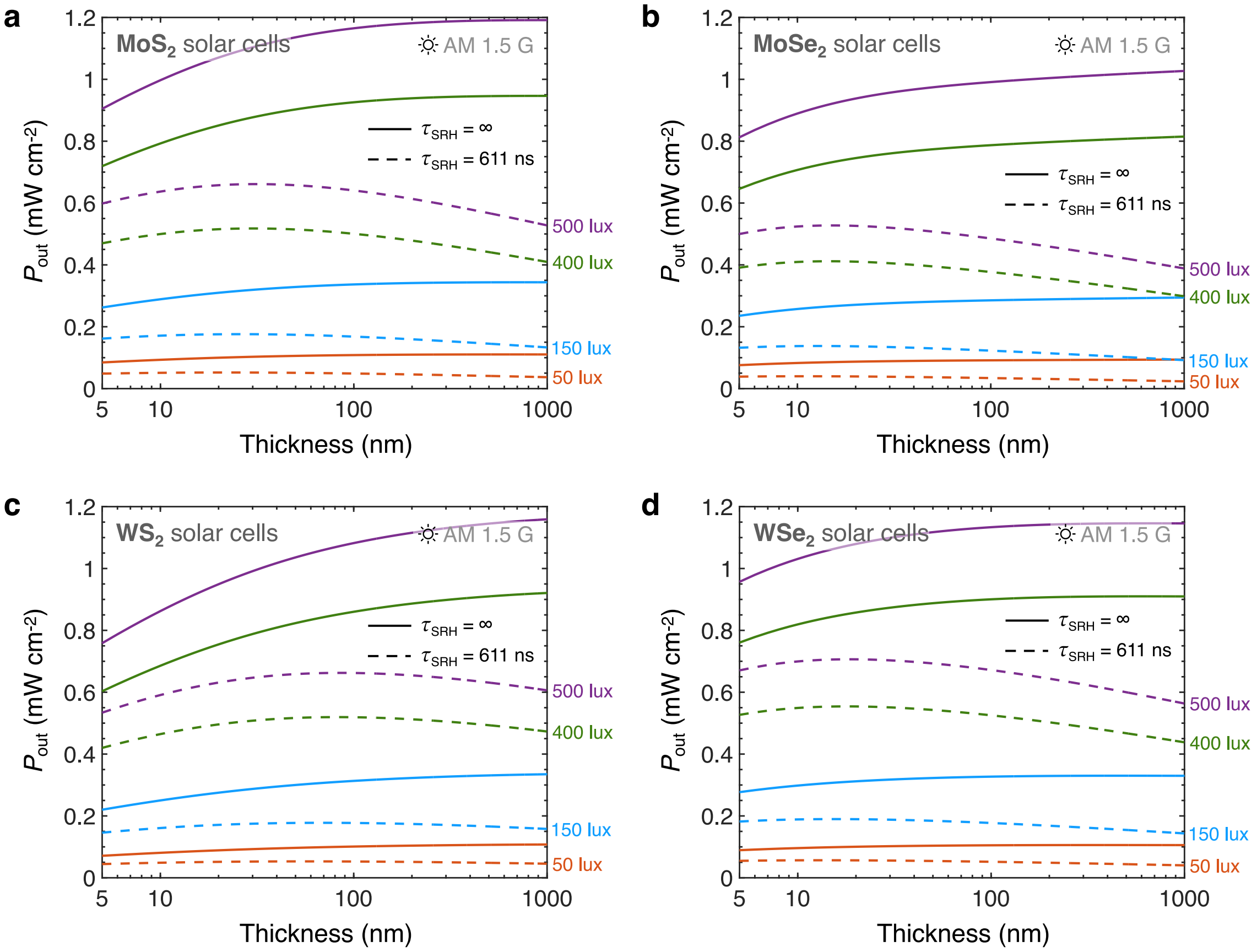

**Supplementary Figure 17 | Output power ($P_{out}$) of thin-film TMD solar cells under AM 1.5 G illumination.** $P_{out}$ of **a,** $MoS_2$, **b,** $MoSe_2$, **c,** $WS_2$, and **d,** $WSe_2$ cells as a function of TMD film thickness, material quality ($\tau_{SRH}$), and AM 1.5 G illumination intensity at 300 K. $\tau_{SRH}$, Shockley-Read-Hall (SRH) lifetime. Solid lines are in the limit defect-free material (no SRH recombination), dashed lines with $\tau_{SRH}$ = 611 ns. Four CFL illumination intensities correspond to the four colors, as labeled (e.g. purple dashed and solid are at 500 lux). Similar to halogen illumination, for an infinite SRH lifetime, $P_{out}$ for all materials consistently rises without a peak, signifying that the positive effects of increased $J_{SC}$ with thickness (**Supplementary Figure 14**) outweigh the negative impacts on $V_{OC}$ (**Supplementary Figure 15**) and FF (**Supplementary Figure 16**). However, like CFL and LED illuminations, when $\tau_{SRH}$ is finite at 611 ns, peaks in $P_{out}$ are evident for all materials. Lower light intensities exhibit flatter peaks, as the gains in $J_{SC}$ under these conditions fail to fully compensate for the associated losses in $V_{OC}$ and FF.

**Supplementary Table 2 | Literature reports on indoor photovoltaic devices.** This table lists power conversion efficiencies (PCE) and associated band gaps for various indoor photovoltaic devices reported in the literature. Band gaps listed with a superscripted, bracketed reference are taken from the referenced study, otherwise they are from the source in the leftmost "Reference" column. Band gaps with an asterisk ('*') denote these were determined using the Tauc method. '**' indicates that the lux intensity of the spectrum was not specified in the reference, and no spectrum was provided, leading us to apply our spectrum to calculate the lux. Note that, even at consistent lux levels and using the same spectra, reported values vary due to differences in indoor spectra across studies.

| **Reference** | **Year** | **Technology** | **Light Source** | **Light Intensity (lux / µW cm$^{-2}$)** | **Band Gap (eV)** | **PCE (%)** |
|---|---|---|---|---|---|---|
| Kao et al.[43] | 2017 | Amorphous Si (a-Si:H) | CFL | 500 / 162 | 1.70 | 12.69 |
| Rossi et al.[44] | 2015 | Single-crystal Si (c-Si) | CFL | 500 / 156.96 | 1.12[45] | 6.11 |
| Liu et al.[46] | 2016 | Dye-sensitized (DSSC) | CFL | 600 / 188.1 | 1.66[47] | 16.1 |
| Freitag et al.[48] | 2017 | Dye-sensitized (DSSC) | CFL | 1000 / 306.6 | 1.93[47] | 28.9 |
| Cao et al.[49] | 2018 | Dye-sensitized (DSSC) | CFL | 500 / 159.1 | 1.50 | 30.8 |
| Michaels et al.[5] | 2020 | Dye-sensitized (DSSC) | CFL | 500 / 151.5 | 1.89 | 32.7 |
| Zhang et al.[50] | 2021 | Dye-sensitized (DSSC) | CFL | 500 / 159.1 | 1.72 | 32.3 |
| Freunek et al.[23] | 2013 | CdTe (II-IV) | CFL | 314.44** / 910 | 1.44 | 10.9 |
| Freitag et al.[48] | 2017 | GaAs (III-V) | CFL | 1000 / 354.0 | 1.42 | 21.0 |
| Antunez et al.[51] | 2017 | CZTSSe (Kesterite) | CFL | 500 / 75 | 1.34 | 11.89 |
| Lee et al.[52] | 2016 | Organic | CFL | 300 / 13.9 | 1.90 | 16.6 |
| Ding et al.[53] | 2019 | Organic | CFL | 1000 / 345 | 1.93 | 26.2 |
| Li et al.[54] | 2018 | Perovskite | CFL | 1000 / 278.7 | 1.55* | 35.2 |
| Li et al.[8] | 2020 | Perovskite | CFL | 1000 / 286.6 | 1.75 | 32.7 |
| Reich et al.[55] | 2011 | Amorphous Si (a-Si:H) | LED | 1000 / 371 | 1.75[45] | 21.0 |
| Kim et al.[56] | 2020 | Amorphous Si (a-Si:H) | LED | 1000 / 310 | 1.75[45] | 29.9 |
| Rossi et al.[44] | 2015 | Single-crystal Si (c-Si) | LED | 500 / 164.9 | 1.12[45] | 4.73 |
| Liu et al.[46] | 2016 | Dye-sensitized (DSSC) | LED | 600 / 179.2 | 1.66[47] | 17.5 |
| Tanaka et al.[57] | 2019 | Dye-sensitized (DSSC) | LED | 1000 / 303.1 | 1.90[47] | 29.2 |
| Teran et al.[58] | 2015 | GaAs (III-V) | LED | 580 / 159.5 | 1.42 | 19.4 |
| Teran et al.[58] | 2015 | $Al_{0.2}Ga_{0.8}As$ (III-V) | LED | 580 / 159.5 | 1.67 | 21.1 |
| Ding et al.[53] | 2019 | Organic | LED | 1000 / 360 | 1.93 | 21.7 |
| Zhang et al.[59] | 2022 | Organic | LED | 500 / 156 | 1.72 | 28.3 |
| Lee et al.[60] | 2023 | Organic | LED | 1000 / 280 | 1.57 | 16.35 |
| Wang et al.[61] | 2023 | Organic | LED | 500 / 157.78 | 1.63* | 29.0 |
| Li et al.[8] | 2020 | Perovskite | LED | 1000 / 279.6 | 1.75 | 35.6 |
| He et al.[62] | 2021 | Perovskite | LED | 824.5 / 301.6 | 1.59 | 40.1 |
| Chen et al.[63] | 2021 | Perovskite | LED | 1000 / 278.8 | 1.54 | 40.99 |
| Gong et al.[9] | 2022 | Perovskite | LED | 1000 / 325 | 1.53* | 41.23 |

## SUPPLEMENTARY REFERENCES

37. Tiedje, T. O. M., Yablonovitch, E. L. I., Cody, G. D. & Brooks, B. G. Limiting efficiency of silicon solar cells. *IEEE Trans Electron Devices* 31, 711–716 (1984).
38. van Bommel, W. Halogen Lamp. in *Encyclopedia of Color Science and Technology* 865–872 (Springer, 2023).
39. Rühle, K. & Kasemann, M. Approaching high efficiency wide range silicon solar cells. in *2013 IEEE 39th Photovoltaic Specialists Conference (PVSC)* 2651–2654 (2013).
40. Young Hugh D. and Freedman, R. A. and F. L. A. and S. F. W. 1898-1975. *University Physics with Modern Physics*. (Pearson Higher Education, Hoboken, N.J., 2020).
41. Jariwala, D. *et al.* Near-unity absorption in van der Waals semiconductors for ultrathin optoelectronics. *Nano Lett* 16, 5482–5487 (2016).
42. Huang, L. *et al.* Atomically thin $MoS_2$ narrowband and broadband light superabsorbers. *ACS Nano* 10, 7493–7499 (2016).
43. Kao, M.-H. *et al.* Low-Temperature Growth of Hydrogenated Amorphous Silicon Carbide Solar Cell by Inductively Coupled Plasma Deposition Toward High Conversion Efficiency in Indoor Lighting. *Sci Rep* 7, 12706 (2017).
44. De Rossi, F., Pontecorvo, T. & Brown, T. M. Characterization of photovoltaic devices for indoor light harvesting and customization of flexible dye solar cells to deliver superior efficiency under artificial lighting. *Appl Energy* 156, 413–422 (2015).
45. Berwal, A. K., Kumari, N., Kaur, I., Kumar, S. & Haleem, A. Investigating the effect of spectral variations on the performance of monocrystalline, polycrystalline and amorphous silicon solar cells. *Journal of Alternate Energy Sources and Technologies* 7, 28–36p (2016).
46. Liu, Y.-C. *et al.* A feasible scalable porphyrin dye for dye-sensitized solar cells under one sun and dim light environments. *J Mater Chem A Mater* 4, 11878–11887 (2016).
47. Asghar, M. I. *et al.* Review of stability for advanced dye solar cells. *Energy Environ Sci* 3, 418–426 (2010).
48. Freitag, M. *et al.* Dye-sensitized solar cells for efficient power generation under ambient lighting. *Nat Photonics* 11, 372–378 (2017).
49. Cao, Y., Liu, Y., Zakeeruddin, S. M., Hagfeldt, A. & Grätzel, M. Direct Contact of Selective Charge Extraction Layers Enables High-Efficiency Molecular Photovoltaics. *Joule* 2, 1108–1117 (2018).
50. Zhang, D. *et al.* A molecular photosensitizer achieves a $V_{oc}$ of 1.24 V enabling highly efficient and stable dye-sensitized solar cells with copper(II/I)-based electrolyte. *Nat Commun* 12, 1777 (2021).
51. Antunez, P. D., Bishop, D. M., Luo, Y. & Haight, R. Efficient kesterite solar cells with high open-circuit voltage for applications in powering distributed devices. *Nat Energy* 2, 884–890 (2017).
52. Lee, H. K. H., Li, Z., Durrant, J. R. & Tsoi, W. C. Is organic photovoltaics promising for indoor applications? *Appl Phys Lett* 108, (2016).
53. Ding, Z., Zhao, R., Yu, Y. & Liu, J. All-polymer indoor photovoltaics with high open-circuit voltage. *J. Mater. Chem. A* 7, 26533–26539 (2019).
54. Li, M. *et al.* Interface modification by ionic liquid: a promising candidate for indoor light harvesting and stability improvement of planar perovskite solar cells. *Adv Energy Mater* 8, 1801509 (2018).

55. Reich, N. H., van Sark, W. G. J. H. M. & Turkenburg, W. C. Charge yield potential of indoor-operated solar cells incorporated into Product Integrated Photovoltaic (PIPV). *Renew Energy* 36, 642–647 (2011).
56. Kim, G., Lim, J. W., Kim, J., Yun, S. J. & Park, M. A. Transparent Thin-Film Silicon Solar Cells for Indoor Light Harvesting with Conversion Efficiencies of 36% without Photodegradation. *ACS Appl Mater Interfaces* 12, 27122–27130 (2020).
57. Tanaka, E., Michaels, H., Freitag, M. & Robertson, N. Synergy of co-sensitizers in a copper bipyridyl redox system for efficient and cost-effective dye-sensitized solar cells in solar and ambient light. *J Mater Chem A Mater* 8, 1279–1287 (2020).
58. Teran, A. S. *et al.* AlGaAs Photovoltaics for Indoor Energy Harvesting in mm-Scale Wireless Sensor Nodes. *IEEE Trans Electron Devices* 62, 2170–2175 (2015).
59. Zhang, T. *et al.* A Medium-Bandgap Nonfullerene Acceptor Enabling Organic Photovoltaic Cells with 30% Efficiency under Indoor Artificial Light. *Advanced Materials* 34, 2207009 (2022).
60. Lee, Y., Biswas, S., Choi, H. & Kim, H. Record High Efficiency Achievement under LED Light in Low Bandgap Donor-Based Organic Solar Cell through Optimal Design. *Int J Energy Res* 2023, 3823460 (2023).
61. Wang, W. *et al.* High-performance organic photovoltaic cells under indoor lighting enabled by suppressing energetic disorders. *Joule* 7, 1067–1079 (2023).
62. He, X. *et al.* 40.1% record low-light solar-cell efficiency by holistic trap-passivation using micrometer-thick perovskite film. *Advanced Materials* 33, 2100770 (2021).
63. Chen, C.-H. *et al.* Ternary two-step sequential deposition induced perovskite orientational crystallization for high-performance photovoltaic devices. *Adv Energy Mater* 11, 2101538 (2021).